\documentclass{article}
\usepackage[utf8]{inputenc}
\usepackage{amsmath, amsthm, amssymb, amscd, amsxtra}
\usepackage[mathscr]{eucal}
\usepackage[all]{xy}
\usepackage{color}
\usepackage{authblk}
\usepackage{tikz}
\usetikzlibrary{decorations.markings}
\usepackage{arydshln}
\usepackage{array, caption, romannum}
\usepackage{verbatim}
\usepackage{enumitem}

\theoremstyle{definition}
\newtheorem{theorem}{Theorem}[section]
\newtheorem{proposition}[theorem]{Proposition}

\newtheorem{lemma}[theorem]{Lemma}
\newtheorem{lemma-definition}[theorem]{Lemma-Definition}

\newtheorem{definition}[theorem]{Definition}
\newtheorem{remark}[theorem]{Remark}

\newcommand{\End}{\operatorname{End}}

\newcommand{\ket}[1]{|#1\rangle}

\title{Ribbon operators in the generalized Kitaev quantum double model based on Hopf algebras}
\author[2]{Bowen Yan \thanks{The first two authors contributed equally to this work.}}
\author[2]{Penghua Chen $ ^*$}
\author[1,2]{Shawn X. Cui \thanks{Corresponding author}}
\affil[1]{{\small Department of Mathematics, Purdue University, West Lafayette}}
\affil[2]{{\small Department of Physics and Astronomy, Purdue University, West Lafayette}}
\affil[ ]{{\small \it \{yan312, chen3014, cui177\} @purdue.edu}}
\date{}

\begin{document}
\maketitle
\pagenumbering{arabic}
\begin{abstract}
Kitaev's quantum double model is a family of exactly solvable lattice models that realize two dimensional topological phases of matter. The model was originally based on finite groups, and was later generalized to semi-simple Hopf algebras. We rigorously define and study ribbon operators in the generalized quantum double model. These ribbon operators are important tools to understand quasi-particle excitations. It turns out that there are some subtleties in defining the operators in contrast to what one would naively think of. In particular, one has to distinguish two classes of ribbons which we call locally clockwise and locally counterclockwise ribbons. Moreover, we point out that the issue already exists in the original model based on finite non-Abelian groups, but it seems to not have been noticed in the literature.  We show how certain common properties would fail even in the original model if we were not to distinguish these two classes of ribbons. Perhaps not surprisingly, under the new definitions ribbon operators satisfy all properties that are expected. For instance, they create quasi-particle excitations only at the end of the ribbon, and the types of the quasi-particles correspond to irreducible representations of the Drinfeld double of the input Hopf algebra. However, the proofs of these properties are much more complicated than those in the case of finite groups. This is partly due to the complications in dealing with general Hopf algebras rather than group algebras.
\end{abstract}

\section{Introduction}
Topological phases of matter (TPM) in two spacial dimensions are gapped quantum liquids at low temperature that have robust ground state degeneracy, stable long-range entanglement, quasi-particle excitations (aka anyons), and possibly non-Abelian exchanging statistics.  There exist global degrees of freedom encoded in the ground states which are resistant to local perturbations and which can be changed unitarily by non-trivial movements of quasi-particle excitations. These features make TPMs ideal quantum media to perform fault-tolerant quantum computing, namely, topological quantum computing \cite{kitaev2003fault} \cite{freedman2002modular}. The theory of TPMs in 2D can be  described equivalently by either a $(2+1)$ topological quantum field theory or a unitary modular tensor category.

A large class of TPMs in 2D are realized by spin lattice models. Among the most well known lies the toric code which is an Abelian toplogical phase and can also be described by a $\mathbb{Z}_2$ gauge theory. Toric code is a special example of Kitaev's quantum double models that associate to each finite group $G$ an exactly solvable lattice model \cite{kitaev2003fault}. When $G$ is $\mathbb{Z}_2$, the theory reduces to toric code. When $G$ is a non-Abelian group, the model realizes a non-Abelian topological phase. In such models, anyon types correspond to irreducible representations of the Hopf algebra $D(G)$, the Drinfeld double (or, quantum double) of the group algebra $\mathbb{C}[G]$.  The quantum double model can be generalized by replacing $G$ with a semi-simple $\mathbb{C}^*$ Hopf algebra $H$. Given such a Hopf algebra, the authors in \cite{buerschaper2013hierarchy} wrote down a frustration-free Hamiltonian consisting of pairwise commuting local projectors analogous to the original setup. We call this model the generalized Kitaev quantum double model\footnote{This model can be further generalized to a semi-simple weak Hopf algebra \cite{chang2014kitaev}. We will not discuss this generalization in this paper.}.

Another class of realizations are the Levin-Wen string-net models \cite{levin2005string} based on unitary fusion categories. String-net models and the quantum double models are closely related. Specifically, for a Hopf algebra $H$, it was shown that the generalized quantum double model based on $H$ is equivalent to the string-net model based on $\text{Rep}(H)$, the category of representations of $H$ \cite{buerschaper2009mapping} \cite{buerschaper2013electric}.

A key tool to describe the creation/annihilation and movement of anyons in the models mentioned above is the notion of ribbon operators (or string operators). In toric code, these are a string of Pauli $Z$ operators on the lattice or a string of Pauli $X$ operators on the dual lattice. However, when the group $G$ is non-Abelian, these two types of string operators have to be `entangled'; one has to consider a thickened string of operators, namely, operators on a ribbon. Roughly, a ribbon is a strip in the lattice with one side running along edges of the lattice and the other side along edges of the dual lattice. In the quantum double model, one first defines the operators for two types of elementary ribbons (triangles), and then extend the definition to longer ribbons using an induction (see \cite{kitaev2003fault} \cite{bombin2008family} for details). In \cite{buerschaper2013hierarchy}, it was stated briefly without proofs that ribbon operators in the generalized quantum double model can also be defined in a similar way. 

In this paper, we rigorously define ribbon operators in the generalized quantum double model based on a semi-simple $\mathbb{C}^*$ Hopf algebra, and systemically study their properties. It is illustrated that the ribbon operators can be interpreted as representations of $D(H)^{*}$ or $D(H)^{*,\text{op}}$, where $D(H)$ is the Drinfeld double of $H$. We also prove explicitly that, given a ribbon, the ribbon operators on it commute with all terms in the Hamiltonian except for those associated with the two ends of the ribbon. Hence, ribbon operators create excitations only at their ends. For a ribbon $\tau$, denote by $V_{\tau}$ the space of states obtained by ribbon operators on $\tau$ acting on the ground state. $V_{\tau}$ is the space of 2-point excitations where the excitations lie at the ends of $\tau$. It is shown that $V_{\tau}$ is naturally isomorphic to $D(H)^*$. Moreover, local operators at the ends of $\tau$ act on $V_{\tau}$ by regular representations of $D(H)$.  It follows that elementary excitation types are in one-to-one correspondence to irreducible representations of $D(H)$. Although these properties are as anticipated, and hence may not be surprising to experts, the computations involved in proving them turn out to be significantly more complicated than those in the case of finite groups. This is partly due to the complications in dealing with general Hopf algebras rather than just group algebras.

Furthermore, we reveal some subtleties in the definition of ribbon operators. In the literature (e.g., \cite{kitaev2003fault}, \cite{bombin2008family}), only two types of elementary ribbons are considered, the direct triangle and the dual triangle. For instance, in Figure \ref{tab:ribbon operators_few}, \Romannum{1} and \Romannum{3} are direct triangles, while \Romannum{2} and \Romannum{4} are dual triangles. However, we show in Section \ref{subsec:definition of ribbon operators} that \Romannum{1} and \Romannum{3} have to be treated differently when defining operators on them, and so do \Romannum{2} and \Romannum{4}. The point is that there is a property, which we call {\it local orientation}, that distinguishes each pair of the above triangles. For instance, \Romannum{2} is locally clockwise while \Romannum{4} is locally counterclockwise. Local orientation can also be extended to general ribbons. As a consequence, there will be two types of ribbons according to their local orientation, and the definition of ribbon operators on each type has to be different. If we were not to distinguish these two types of ribbons, certain common properties to be expected would not hold. For example, the ribbon operator would fail to commute with terms of the Hamiltonian away from the end points. Surprisingly, we point out that this issue already exists even in the original quantum double model when the input group is non-Abelian, but this issue seems to not have been addressed in the literature to the best of our knowledge.  Lastly, our definition of ribbon operators is explicit, in contrast to those in the string-net models where one needs to solve a set of consistency equations.  
\begin{table}
\begin{tabular}{m{11cm}}
\centering
\begin{tikzpicture}
\draw[step=2, dashed, gray] (-7.5,-3.5) grid (3.5,1.5);
\draw[step=2, gray, shift={(1,1)}] (-8.5,-4.5) grid (2.5,0.5);
\begin{scope}[xshift = -2cm]
\draw[black, thick] (0,0)--(-1,1)--(-1,-1)--cycle;
\draw[black, thick] (1,1)--(0,0)--(2,0)--cycle;
\draw[black, thick] (0,-2)--(-1,-1)--(-1,-3)--cycle;
\draw[black, thick] (1,-3)--(0,-2)--(2,-2)--cycle;
\draw[double, ->, >=stealth, black] (-0.65,-0.3)--(-0.65,0.3) node[anchor=north west] {\Romannum{1}};
\draw[double, ->, >=stealth, black] (0.7,0.35)--(1.3,0.35) node[anchor=south east] {\Romannum{2}};
\draw[double, ->, >=stealth, black] (-0.65,-1.7)--(-0.65,-2.3) node[anchor=south west] {\Romannum{3}};
\draw[double, ->, >=stealth, black] (0.7,-2.35)--(1.3,-2.35) node[anchor=north east] {\Romannum{4}};
\draw[very thick] (-1,-1)--(0,0);
\end{scope}
\end{tikzpicture}
\end{tabular}

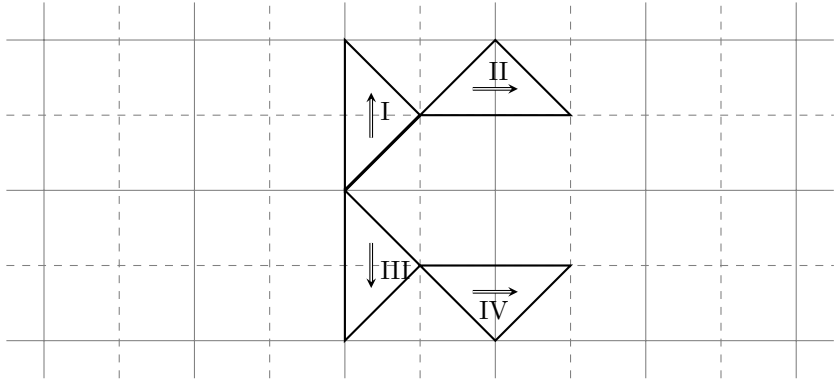
\captionof{figure}[foo]{An illustration of elementary ribbons (dark solid triangles). The solid grid represents the lattice and the dashed grid represents the dual lattice.}
\label{tab:ribbon operators_few}
\end{table}

The rest of the paper is organized as follows. In Section \ref{sec:background}, we collect some basic facts about semi-simple Hopf algebras, their representations, and Drinfeld double. We also review the Hamiltonian of the generalized quantum double model. In Section \ref{sec:ribbon_operators_all}, we carefully formulate ribbons, provide the definition of ribbon operators, and study their properties. In particular, it is shown in Section \ref{subsec:local orientation original} that local orientation needs to be considered even in the original Kitaev model with a non-Abelian group. Many of technical details can be found in the appendices.

\section{Background}\label{sec:background}
\subsection{Hopf algebra} \label{sec:hopf algebra}
Hopf algebras are important objects in a number of areas, such as representation theory, tensor categories, algebraic topology, topological quantum field theories, etc. There is an extensive literature covering different aspects of Hopf algebras.  In this section, we simply provide a brief review with the main purpose of fixing conventions. For detailed discussions, see for instance \cite{radford2011hopf} \cite{kassel2012quantum}.

A Hopf algebra over $\mathbb{C}$ is a vector space $H$ endowed with the linear maps (called structure maps),
\begin{align}
&\mu\colon H \otimes H \to H, \quad&\eta\colon \mathbb{C} \to H,\\
&\Delta\colon H \to H \otimes H, \quad&\epsilon\colon H\to \mathbb{C},\\
&S\colon H \to H,
\end{align}
satisfying several conditions to be specified in the following. 

Firstly, $(\mu, \eta)$ defines an (associative) algebra structure. That is, the multiplication $\mu$ is associative:
\begin{equation}
    \mu\left[\mu (a\otimes b)\otimes c\right]=\mu\left[a\otimes \mu(b\otimes c)\right],
\end{equation}
or briefly
\begin{equation}
    (ab)c=a(bc).
\end{equation}
The unit $1_H$ for the multiplication $\mu$ is given by $\eta(1)$. Secondly, $(\Delta, \epsilon)$ defines a (coassociative) coalgebra structure with $\Delta$ and $\epsilon$ the comultiplication and counit, respectively. We will use the {\it Sweedler notation} for expressions involving comultiplications. For instance, we write
\begin{equation} \label{eqn:delta}
    \Delta(a)=\sum_{(a)} a'\otimes a''.
\end{equation}
The comultiplication map being coassociative means
\begin{equation}
    (\Delta \otimes id) \circ \Delta=(id \otimes \Delta) \circ \Delta,
\end{equation}
or in Sweedler notation,
\begin{equation}
    \sum_{(a)} \left[\sum_{(a')} (a')' \otimes (a')''\right]\otimes a''= \sum_{(a)} a'\otimes \left[\sum_{(a'')} (a'')' \otimes (a'')''\right].
\end{equation}
Due to the above equality, we simply write
\begin{equation}
    (\Delta \otimes id) \circ \Delta (a) = \sum_{(a)} a' \otimes a'' \otimes a''', 
\end{equation}
or
\begin{equation}
    (\Delta \otimes id) \circ \Delta (a) = \sum_{(a)} a^{(1)} \otimes a^{(2)} \otimes a^{(3)}.
\end{equation}
More generally, we use the Sweedler notation for
\begin{equation}
   (\Delta \otimes id_{H^{\otimes (n-2)}}) \circ\cdots\circ(\Delta \otimes id) \circ \Delta (a) = \sum_{(a)} a^{(1)} \otimes \cdots \otimes a^{(n)}.
\end{equation}
The counit $\epsilon$ satisfies
\begin{equation}
    \sum_{(a)} \epsilon(a')a''=\sum_{(a)} a'\epsilon(a'')=a.
\end{equation}
Thirdly, $\Delta$ and $\epsilon$ are both required to be  algebra morphisms. In particular, this implies $\epsilon$ defines a 1-dimensional representation of $H$. Lastly, 
$S$ is called the antipode which is invertible in our consideration satisfying:
\begin{equation}
    \sum_{(a)} a' S(a'') = \epsilon(a) 1_H = \sum_{(a)} S(a') a''.
\end{equation}
To emphasize on structure maps, we also denote a Hopf algebra by
\begin{equation}
    (H;\mu,\eta,\Delta,\epsilon,S).
\end{equation}

In this paper, we will only consider finite dimensional semisimple Hopf algebras. Over $\mathbb{C}$, semisimplicity is equivalent to the condition that $S$ is involutory, namely, $S^2 = id$. The following identities are implied in a finite dimensional Hopf algebra. 
\begin{equation}
    S(ab) = S(b)S(a),\quad S(1_H)=1_H,\quad
    \epsilon[S(a)] = \epsilon(a),
\end{equation}
\begin{equation}
    \sum_{(a)} S(a'') \otimes S(a') = \sum_{(S(a))} S(a)' \otimes S(a)''.
\end{equation}

Given a Hopf algebra $(H;\mu,\eta,\Delta,\epsilon,S)$, there are several ways of constructing new Hopf algebras out of it. Take $H^*$ to be the linear dual of $H$. Then 
\begin{equation}
    ( H^*; \Delta^T,\epsilon^T,\mu^T,\eta^T, S^T),
\end{equation}
defines a Hopf algebra structure on $H^*$, where for a map $f$, $f^T$ means the  linear dual of $f$.\footnote{Another common notation for $f^T$ is $f^*$. Here we use $f^T$ since under appropriate bases, the matrix of $f^T$ is the transpose of that of $f$. Another reason is to avoid confusion since we will introduce a $*$ operation below with a different meaning.} For example, $\mu^T$ is a map from $H^*$ to $H^* \otimes H^*$: 
\begin{equation}
    \mu^T (f) (a \otimes b) = f \left[\mu(a \otimes b)\right] = f(ab).
\end{equation}
where $a,b \in H$, and $ f \in H^* $. We can also define the opposite Hopf algebra $H^{op}$ by
\begin{equation}
    (H^{op};\mu^{op},\eta, \Delta,\epsilon, S^{-1}),
\end{equation}
where $H^{op}$ as a vector space is the same as $H$, and $\mu^{op}$ is defined as
\begin{equation}
    \mu^{op}(a \otimes b)= \mu(b \otimes a) = ba.
\end{equation}
Similarly, we have the co-opposite Hopf algebra $H^{cop}$,
\begin{equation} \label{Def:H_cop}
   (H^{cop};\mu,\eta, \Delta^{cop},\epsilon, S^{-1}),
\end{equation}
where again $H^{cop}$ as a vector space is $H$, and   $\Delta^{cop}$ is defined as
\begin{equation} \label{eqn:cop of Delta}
    \Delta^{cop}(a)= \sum_{(a)} a'' \otimes a'.
\end{equation}
The above three operations $(\cdot)^*, \ (\cdot)^{op},$ and $(\cdot)^{cop}$ are all involutive, and can also be composed with each other. It is direct to check that as Hopf algebras $(H^*)^{cop} \simeq (H^{op})^*$ and $(H^*)^{op} \simeq (H^{cop})^*$.

For a semisimple Hopf algebra $H$, a (two-sided) integral is an element $h_0 \in H$ such that for all $a \in H$,
\begin{equation}\label{equ:h_0 def1}
    ah_0=h_0a=\epsilon(a)h_0.
\end{equation}
The space of integrals is  a 1-dimensional subspace, and hence $h_0$ is uniquely defined if we require
\begin{equation}\label{equ:h_0 normalization}
    h_0^2=h_0, \quad \text{ or equivalently } \quad \epsilon(h_0) = 1.
\end{equation}
We call such an $h_0$ the Haar integral of $H$.  It can be proved that $h_0$ is cocommutative, namely
\begin{equation}
    \Delta(h_0)=\sum_{(h_0)} h_0' \otimes h_0'' = \sum_{(h_0)} h_0'' \otimes h_0'.
\end{equation}

To make a Hopf algebra into a Hilbert space, we introduce the $*$-structure. A $*$-structure on $H$ is a conjugate-linear map $*: H \to H$ satisfying the following properties:
\begin{equation} \label{eqn: properties of *}
    (a^*)^{*}=a,\quad (ab)^{*}=b^*a^*,\quad 1^{*}=1,
\end{equation}
\begin{equation}
    \sum_{(a)} (a')^* \otimes (a'')^*=\sum_{(a^*)} (a^*)' \otimes (a^*)''.
\end{equation}
A Hopf algebra endowed with a $*$-structure is called a $\mathbb{C}^*$ Hopf  algebra. Let $H$ be such a Hopf algebra, and denote by $\phi$  the
Haar integral of $H^*$. For $a,b \in H$, define 
\begin{equation} \label{eqn:the relation bewteen trace and inner product}
    \langle a,b \rangle = \phi(a^*b).
\end{equation}
The above form $ \langle \cdot,\cdot \rangle$ defines a Hermitian inner product on $H$.

Unless otherwise stated, throughout this paper we will,  as a convention, use letters such as $h_0$, $\phi$ for the Haar integrals,  $a$, $b$, $c$, $x$, $y$ for general elements of $H$, and  $f$, $g$, $t$ for general elements of $H^*$. We adopt the notation that $f(x?)$ is an element of $H^*$ such that $f(x?)(y) = f(xy)$.

\subsection{Representations of semisimple Hopf algebras}\label{subsec:rep of Hopf}
The category of finite dimensional representations over $\mathbb{C}$ of a semisimple Hopf algebra $H$ is a semisimple tensor category with duals. If $V, \ W$ are two representations,
\begin{align}
    \rho_V\colon &H \to \End(V),\\
    \rho_W\colon &H \to \End(W),
\end{align}
then $V \otimes W$ is a representation with the action given by,
\begin{align}
    a.(v \otimes w):= \bigl((\rho_V \otimes \rho_W)\Delta(a)\bigr)(v \otimes w), \quad a \in H, v \in V, w \in W,
\end{align}
and so is $V^*$ with the action given by,
\begin{align}
    a. f := f \circ \rho_V(S(a)), \quad a \in H, f \in V^*.
\end{align}

A representation $V$ of $H$ is irreducible if $\End_{H}(V) \simeq \mathbb{C}$. Denote by $\text{Irr}_H$ the set of isomorphism classes of irreducible representations of $H$. Consider the  regular representation $H$ with the action given by left multiplication,
\begin{equation}
    L(a)(c):=ac,
\end{equation}
or by right multiplication by $S(\cdot)$,
\begin{equation}
    R(a)(c):=cS(a).
\end{equation}
These two actions commute and hence define an action of $H \otimes H$ on $H$ by,
\begin{align}
    (a \otimes b). c := acS(b).
\end{align}
It is a basic fact that as a representation of $H \otimes H$, we have the isomorphism,
\begin{align}\label{equ:H_regular_decomposition}
    H \simeq \bigoplus_{\mu \in \text{Irr}_H} \mu^* \otimes \mu.
\end{align}
An explicit isomorphism is given as follows. For each $\mu \in \text{Irr}_H$, fix a basis $\{| i \rangle \ | \ i = 1, \cdots, dim(\mu)\}$, and denote the matrix of an element $a \in H$ under this basis by $D^{\mu}(a)$. Let $h_0 \in H$ be the Haar integral (see Equations \ref{equ:h_0 def1}, \ref{equ:h_0 normalization}). We define the `Fourier transformation' on $H$ by \cite{buerschaper2013electric},
\begin{equation}
    |\nu ij\rangle=\sqrt{\frac{dim(\nu)}{dim(H)}} \sum_{(h_0)} D^{\nu} (h_0')_{ij}h_0'',
\end{equation}
where $\nu \in \text{Irr}_H$, and  $i,j = 1,2,\cdots, dim(\nu)$. For self-containedness, in Appendix \ref{sec:FT of $H^*$}, we verify that the action of $H \otimes H$ on the subspace $\text{span}\{|\nu ij\rangle\ |\ i,j = 1, \cdots, dim(\nu)\}$ is given by $\nu^* \otimes \nu$, and hence defines a desired isomorphism for Equation \ref{equ:H_regular_decomposition}.

Lastly, the two representations $L$ and $R$ each induce a representation of $H$ on $H^*$,
\begin{equation} \label{eqn:left action on H^*}
    L(a)|f\rangle=|f[S(a)?]\rangle,
\end{equation}
\begin{equation} \label{eqn:right action on H^*}
    R(a)|f\rangle=|f(?a)\rangle, \quad \ket{f} \in H^*.
\end{equation}

\subsection{Drinfeld double of Hopf algebras}
The Drinfeld double (or quantum double) $D(H)$ of a Hopf algebra $H$ is a Hopf algebra
\begin{equation} \label{Def:Quantum Double 1}
    D(H) = \bigl((H^*)^{cop} \otimes H; \mu_D, \eta_D,\Delta_D,\epsilon_D,S_D\bigr).
\end{equation}
It is constructed as a bicrossed product of $H$ and $(H^*)^{cop}$. For $f, g \in H^*$, $a, b \in H$, $\mu_D$ is defined as
\begin{equation} \label{eqn:staightening equation}
    \mu_D \left[(f \otimes a) \otimes (g \otimes b) \right] = \sum_{(a)} f \,g\left[S^{-1}(a''')?a' \right] \otimes  a''b,
\end{equation}
which is known as the \textit{straightening equation}. Notice that we have
\begin{equation}
    f \otimes a = (f \otimes 1) (1 \otimes a).
\end{equation}
The other structure maps can be determined by the property that $(H^*)^{cop}$ and $H$ are both sub Hopf algebras of $D(H)$ by the inclusions $f \mapsto f \otimes 1$ and $a \mapsto \epsilon \otimes a$, respectively.  For example, $\Delta_D$ is given by,
\begin{equation}
    \Delta_D (f \otimes a) = \sum_{(f),(a)} (f'' \otimes a') \otimes (f' \otimes a''),
\end{equation}
where in the Sweedler notation of $f$, we treat $f$ as an element of $H^*$ rather than $(H^*)^{cop}$. We will also use this convention throughout the paper. Namely, for  $a \in H^{cop}$, as far as the Sweedler notation is concerned, we use $\Delta$ rather than $\Delta^{cop}$ to define $a', a''$, etc. Other structure maps are provided as follows,
\begin{equation}
    \eta_D(1) = \epsilon \otimes 1,
\end{equation}
\begin{equation}
    \epsilon_D(f \otimes a) = f(1) \otimes \epsilon(a),
\end{equation}
\begin{equation}
    S_D(f \otimes a) = S(a)S^T(f).
\end{equation}

\subsection{Generalized Kitaev model based on Hopf algebras} \label{sec:Kitaev}
In this subsection, $H$ denotes a semisimple $\mathbb{C}^*$ Hopf algebra. The original Kitaev model \cite{kitaev2003fault} is constructed based on the group algebra $\mathbb{C}[G]$ of a finite group $G$ while the generalized Kitaev model is based on a semisimple $\mathbb{C}^*$ Hopf algebra $H$. The latter is introduced in \cite{buerschaper2013hierarchy} which we review below.

For simplicity, we take a square lattice $\Gamma=(V,E,P)$ to establish the model, where $V, \ E,$ and $P$ denote the set of vertices, (directed) edges, and faces, respectively, as shown in Figure \ref{tab:square lattices} (the solid grid)\footnote{The edges in the lattice can be arbitrarily directed, and the physics of the model will be independent of those directions.}. We also define the dual lattice $\Gamma^*=(P^*,E^*,V^*)$ where $P^*$ is the set of vertices in $\Gamma^*$  dual to the faces $P$ in $\Gamma$, and $E^*$ and $V^*$ have similar interpretations. For an element $x \in V \cup E \cup P$, denote by $x^*$ the corresponding element in $V^* \cup E^* \cup P^*$. For an edge $e \in E$, the direction of the dual edge $e^*$  is obtained by rotating the direction of $e$ counterclockwise by $90^{\circ}$.
A \textbf{site} $s=(v, p)$ is a pair of a vertex $v$ and an adjacent face $p$ containing $v$. We draw a segment connecting $v$ and the dual vertex $p*$ to represent the site. See Figure \ref{tab:square lattices}.

To each edge $e$ of $\Gamma$, we attach a copy of the Hopf algebra (also a Hilbert space) $\mathcal{H}_e := H$.  The total Hilbert space of the model is the tensor product over all edges of the associated Hilbert spaces:
\begin{equation}
    \mathcal{H} := \bigotimes_{e \in E} \mathcal{H}_e.
\end{equation}
\begin{equation}\label{equ:edgeL}
    L_+^a (x) = ax,\quad L_-^a (x) = xS(a).
\end{equation}
\begin{equation}\label{equ:edgeT}
    T_+^f (x) = f(x'')x',\quad T_-^f (x) = f[S(x')]x''.
\end{equation}
\begin{table}[ht]
\begin{tabular}{m{11cm}} 
\centering
\begin{tikzpicture}
\draw[step=2, dashed, gray, thick] (-1.3,-1.9) grid (5.6,1.9);
\draw[step=2, gray, thick, shift={(1,1)}] (-2.3,-2.9) grid (4.6,0.9);

\draw[->, black, very thick] (-1,0)--(-1,-0.65);
\draw[->, dashed, black, very thick] (-0.9,0)--(-0.35,0);
\draw[->, gray, very thick] (-1,-0.25) arc (-90:-10:0.25);
\node at (-0.5,-0.4) {$90^{\circ}$};

\filldraw[black] (2,0) circle (1.5pt) node[anchor=south east] {$p^*$};
\filldraw[black] (1,-1) circle (1.5pt) node[anchor=south east] {$v$};
\draw[red, very thick] (1,-1)--(2,0) node[anchor=north west, black, pos = .5] {$s$};
\filldraw[red] (2,0) circle (2pt) node[anchor=north west] {};

\draw[black, very thick] (5,-1)--(5,1);
\draw[->, black, very thick] (5,-0.01)--(5,0);
\node at (5.3,-0.3) {$T_{-}^{f}$};
\node at (4.7,0.3) {$T_{+}^{f}$};
\node at (5.3,1.3) {$L_{-}^{a}$};
\node at (4.7,-1.4) {$L_{+}^{a}$};
\end{tikzpicture}
\end{tabular}

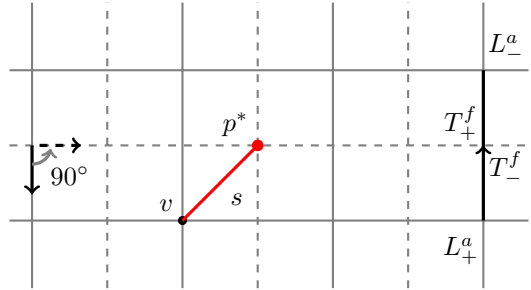
\captionof{figure}[foo]{The solid grid connecting all vertices $V$ represents the square lattice $\Gamma$, while the dashed grid connecting all dual vertices $P^*$ represents the dual square lattice $\Gamma^*$. A site $s=(v, p)$ is represented by a segment connecting a vertex $v$ and a dual vertex $p^*$. For $f \in H^*, a \in H$, the edge operators $T_{\pm}^{f}$ and $L_{\pm}^{a}$ act on the Hilbert space $\mathcal{H}_e$ of an edge $e$.}
\label{tab:square lattices}
\end{table}

Upon the establishment of the oriented graph $\Gamma=(V,E,P)$, we can define the edge operators \cite{buerschaper2013hierarchy}  illustrated in Figure \ref{tab:square lattices} and the local operators $A_a(s)$ and $B_f(s)$ on a site $s=(v , p)$ illustrated in Figure \ref{tab:definition of A} and Figure \ref{tab:definition of B}, respectively. For each edge $e$ of the lattice and for 
$f \in H^*, a \in H$, the edge operators $T_{\pm}^{f}$ and $L_{\pm}^{a}$ act on the Hilbert space $\mathcal{H}_e$. See Equations \ref{equ:edgeL}-\ref{equ:edgeT}.

To define $A_a(s)$ for $a \in H$, we start from the site $s$, go around the vertex $v$ to apply edge operators $L_{\pm}^{a'}$, $L_{\pm}^{a''}$, $L_{\pm}^{a'''}$, $L_{\pm}^{a^{(4)}}$ to each edge adjacent to $v$ in counterclockwise order as shown and explained in Figure \ref{tab:definition of A}. For example, when it is applied to the product state of $|x_1\rangle$, $|x_2\rangle$, $|x_3\rangle$, $|x_4\rangle$ for the configuration in Figure \ref{tab:definition of A}, the result is
\begin{equation}
    A_a(s) |x_1\rangle|x_2\rangle|x_3\rangle|x_4\rangle = \sum |a' x_1\rangle|a'' x_2\rangle|a'''x_3\rangle|a^{(4)} x_4\rangle.
\end{equation}
\begin{equation}
    A_{a}(s) = \sum L_{+}^{a'} \otimes L_{+}^{a''} \otimes L_{+}^{a'''} \otimes L_{+}^{a^{(4)}}
\end{equation}
\begin{table}[ht]
\begin{tabular}{m{11cm}}
\centering
\begin{tikzpicture} 
\draw[black, thick] (-1.5,0)--(1.5,0);
\draw[black, thick] (0,-1.5)--(0,1.5);
\draw[->, black, thick] (0,0)--(0,1) node[anchor=south east] {$x_1$};
\draw[->, black, thick] (0,0)--(-1,0) node[anchor=north east] {$x_2$};
\draw[->, black, thick] (0,0)--(0,-1) node[anchor=north west] {$x_3$};
\draw[->, black, thick] (0,0)--(1,0) node[anchor=south west] {$x_4$};
\filldraw[red] (1,1) circle (2pt);
\draw[red, very thick] (0,0)--(1,1);
\draw[->, black, thick] (0.12941,0.48296) arc (75:375:0.5);
\filldraw[black] (0,0) circle (2pt);
\node[black] at (-0.2,0.2) {$v$};
\node[black] at (0.4,0.7) {$s$};
\end{tikzpicture}
\end{tabular}
\captionof{figure}[foo]{The convention for the local operator $A_{a}(s)$: for each edge, we choose $+$ sign for the edge operator $L^{a^{(n)}}$  if the edge leaves the vertex, and choose $-$ sign otherwise.}
\label{tab:definition of A}
\end{table}

To define $B_f(s)$, $f \in H^*$, we start from the site $s$, go around the dual vertex $p^*$ to apply edge operators $T_{\pm}^{f'}$, $T_{\pm}^{f''}$, $T_{\pm}^{f'''}$, $T_{\pm}^{f^{(4)}}$ to the edges on the boundary of $p$ in counterclockwise order as shown and explained in Figure \ref{tab:definition of B}. When it is applied to the product state of $|x_1\rangle$, $|x_2\rangle$, $|x_3\rangle$, $|x_4\rangle$ for the configuration of Figure \ref{tab:definition of B}, the result is \footnote{To derive Equation \ref{eqn:derivation of B from definition}, we use the fact that the comultiplication $\Delta_*$ in $H^*$ is actually $\mu^T$.}
\begin{equation}\label{eqn:derivation of B from definition}
B_f(s) |x_1\rangle|x_2\rangle|x_3\rangle|x_4\rangle = \sum f(x_1''x_2''x_3''x_4'')|x_1'\rangle|x_2'\rangle|x_3'\rangle|x_4'\rangle.
\end{equation}
\begin{equation}
B_{f}(s) = \sum T_{+}^{f'} \otimes T_{+}^{f''} \otimes T_{+}^{f'''} \otimes T_{+}^{f^{(4)}}
\end{equation}
\begin{table}[ht]
\begin{tabular}{m{10cm}}
\centering
\begin{tikzpicture} 
\draw[black, thick] (-1,-1) rectangle (1,1);
\draw[->, black, thick] (-0.5,-1)--(0,-1) node[anchor=north] {$x_1$};
\draw[->, black, thick] (1,-1)--(1,0) node[anchor=west] {$x_2$};
\draw[->, black, thick] (1,1)--(0,1) node[anchor=south] {$x_3$};
\draw[->, black, thick] (-1,1)--(-1,0) node[anchor=east] {$x_4$};
\filldraw[red] (0,0) circle (2pt);
\draw[red, very thick] (-1,-1)--(0,0);
\draw[->, black, thick] (-0.12941,-0.48296) arc (-105:195:0.5);
\node[black] at (0.3,0) {$p^*$};
\node[black] at (-0.4,-0.7) {$s$};
\end{tikzpicture}
\end{tabular}
\captionof{figure}[foo]{The convention for the local operator $B_{f}(s)$: for each edge, we choose $+$ sign for the edge operator $T^{f^{(n)}}$ if  the direction of the edge coincides with the counterclockwise orientation of the boundary of $p$,  and choose $-$ sign otherwise.}
\label{tab:definition of B}
\end{table}
\begin{remark}
We remark that our convention for defining the operators  $A_a(s)$ and $ B_f(s)$ is opposite to that in \cite{kitaev2003fault} \cite{buerschaper2013hierarchy}. Explicitly, these operators on a lattice $\Gamma$ will be the same as those of  \cite{buerschaper2013hierarchy} on a lattice $\Gamma'$ obtained from $\Gamma$ by reversing the orientation of all edges. When the Hopf algebra is a group algebra, our convention is consistent with that in \cite{bombin2008family}.
\end{remark}

For each site $s = (v,p)$, we extend the definition of $A_a(s),\ B_f(s)$ to the whole Hilbert space $\mathcal{H}$ by tensoring the identity operator on edges not adjacent to $v$ or $p$.  The $A_a(s) $ and $\ B_f(s)$ are called local operators at $s$. They define a representation of the Drinfeld double $D(H)$  by mapping $f \otimes a$ to $B_fA_a$. The most nontrivial part of the statement is the straightening equation. For self containedness, we verify the straightening equation for the local operators in Appendix \ref{sec:Straightening equation of $A_a$ and $B_f$}. 

Let $h_0 \in H$ and $\phi \in H^*$ be the Haar integral. For a site $s = (v, p)$, it can be checked that $A_{h_0}(v):= A_{h_0}(s)$ only depends on $v$ and $B_{\phi}(p):= B_{\phi}(s)$ only depends on $p$. Moreover, the set of operators $\{A_{h_0}(v)\ :\ v \in V\} \cup \{B_{\phi}(p)\ :\ p \in P\}$  are mutually commuting projectors. The (frustration-free) Hamiltonian of the model is given by, 
\begin{equation}
    H = -\sum_{v \in V} A_{h_0}(v) - \sum_{p \in P} B_{\phi}(p).
\end{equation}

The ground states are simultaneously stabilized by all the terms in the Hamiltonian. Equivalently, the ground states space can be characterized as the subspace of $\mathcal{H}$ corresponding to the trivial representation of $D(H)$ on all sites $s$. 

\section{Ribbon operators}\label{sec:ribbon_operators_all}
In this section, we rigorously define ribbons, operators on them called ribbon operators, and study some of their important properties.
\subsection{Directed ribbons}\label{subsec:directed_ribbons}
Let $s_0 = (v_0, p_0)$ and $s_1 = (v_1, p_1)$ be two distinct sites that share a common vertex (i.e., $v_0 = v_1$) or a common dual vertex (i.e., $p_0 = p_1$). There is a unique triangle $\tau$ whose sides are given by $s_0$, $s_1$, and an edge $e_{\tau}$ in the lattice or the dual lattice. See the bottom left two examples in Figure \ref{tab:ribbon operators}. The triangle $\tau$ is said to be of dual (resp. direct) \textbf{type} if $e_{\tau}$ is an edge in the dual (resp. direct) lattice, or equivalently, if $v_0 = v_1$ (resp. $p_0 = p_1$). We also assign a direction to $\tau$, indicated by a double arrow inside the triangle, so that it points from $s_0$ to $s_1$. Denote by $s_i = \partial_i \tau,\ i =0, 1$. A {\it ribbon} is a sequence of mutually non-overlapping directed triangles $\tau = \tau_1\tau_2 \cdots \tau_n$ such that $\partial_1 \tau_i = \partial_0 \tau_{i+1}, \ i = 1, \cdots, n-1$. Note that $\tau$ inherits a direction from its components, also indicated by a double arrow, and we call $\partial_0 \tau := \partial_0 \tau_1$ the initial site and  $\partial_1 \tau := \partial_1 \tau_n$ the terminal site of $\tau$. See Figure \ref{tab:ribbon operators} for an illustration of several ribbons. By default, all ribbons are directed. A closed ribbon is one for which the initial site and terminal site coincide. Unless otherwise stated, ribbons considered in this paper are not closed. Triangles are called elementary ribbons.

We introduce a property,  called \textbf{local orientation}, of directed ribbons which seems to be missing in the literature, but will turn out to be critical to coherently define ribbon operators. 
\begin{definition}
Let $\tau$ be a directed triangle (of dual or direct type) with initial site $s_0 = \partial_0 \tau = (v_0, p_0)$. Then $\tau$ has clockwise (resp. counterclockwise) local orientation if a clockwise (resp. counterclockwise) rotation of $s_0$ around $p_0^*$ immediately swipes through the interior of $\tau$.  We draw a clockwise/counterclockwise arrow around $p_0^*$ to denote the local orientation of $\tau$ (See Figure \ref{tab:ribbon operators}).
\end{definition}

An intuitive motivation for introducing local orientation is as follows. We can see that for a triangle of a given type, a choice of direction is not sufficient to uniquely determine the shape of the triangle. For example, the triangles \Romannum{2} and \Romannum{4} in Figure \ref{tab:ribbon operators} are both of dual type and directed to the right, but \Romannum{4} is an `upside down' version of \Romannum{2}, and as will be shown later, they have to be treated differently when we define ribbon operators on them. Local orientation can be used to distinguish those two since triangle \Romannum{2} is locally clockwise while \Romannum{4} is locally counterclockwise.

It is straightforward to see that changing the direction of a triangle will also change its local orientation. We note that a choice of direction is a {\it structure} on the triangle, while the type and local orientation are each a {\it property} of a directed triangle (though only the later depends on the direction). Thus, there are four classes of  directed triangles according to different combinations of  local orientation and type. In Figure \ref{tab:ribbon operators}, the triangles \Romannum{1}-\Romannum{4} in increasing order are, respectively, clockwise direct, clockwise dual, counterclockwise direct, and counterclockwise dual.

Now let $\tau$ be a general directed ribbon. Clearly, its composite triangles can have different types (direct or dual).  However, an important observation is that all of the triangles of $\tau$ must have the same local orientation. Hence, we can extend the notion of local orientation from triangles to general ribbons. Intuitively, if a ribbon aligns horizontally and directs from left to right, then turning it upside down will change its local orientation while keeping its direction. Reversing the direction alone will  flip its local orientation as well. As a notation, we also denote a directed ribbon by $\tau_L$ if it is locally clockwise and by  $\tau_R$ if it is locally counterclockwise. (This notation is motivated by the left/right hand rule)

\begin{table}
\begin{tabular}{m{11cm}}
\centering
\begin{tikzpicture}
\draw[step=2, dashed, gray] (-7.5,-3.5) grid (3.5,1.5);
\draw[step=2, gray, shift={(1,1)}] (-8.5,-4.5) grid (2.5,0.5);
\draw[black, thick] (-6,0)--(-7,-1)--(-5,-1)--cycle;
\draw[black, thick] (-6,0)--(-4,0)--(-5,-1)--cycle;
\draw[black, thick] (-3,-1)--(-4,0)--(-5,-1)--cycle;
\draw[black, thick] (-3,-1)--(-4,0)--(-2,0)--cycle;
\filldraw[red] (-6,0) circle (2pt);
\draw[red, very thick] (-7,-1)--(-6,0);
\filldraw[blue] (-2,0) circle (2pt);
\draw[blue, very thick] (-3,-1)--(-2,0);
\node[black] at (-6.7,-0.3) {$s_0$};
\node[black] at (-2.3,-0.7) {$s_1$};
\draw[double, ->, >=stealth, black] (-4.7,0.5)--(-3.7,0.5); 
\node at (-4.2,0.7) {$\tau$};
\draw[double, ->, >=stealth, black] (-6.3,-0.65)--(-5.7,-0.65) node[anchor=south east] {$\tau_{1}$};
\draw[double, ->, >=stealth, black] (-5.3,-0.35)--(-4.7,-0.35) node[anchor=north east] {$\tau_{2}$};
\draw[double, ->, >=stealth, black] (-4.3,-0.65)--(-3.7,-0.65) node[anchor=south east] {$\tau_{3}$};
\draw[double, ->, >=stealth, black] (-3.3,-0.35)--(-2.7,-0.35) node[anchor=north east] {$\tau_{4}$};
\draw[black, thick] (-6,-2)--(-7,-3)--(-5,-3)--cycle;
\draw[red, very thick] (-7,-3)--(-6,-2);
\draw[blue, very thick] (-5,-3)--(-6,-2);
\filldraw[purple] (-6,-2) circle (2pt);
\node[black] at (-6.7,-2.3) {$s_0$};
\node[black] at (-5.3,-2.3) {$s_1$};
\draw[double, ->, >=stealth, black] (-6.3,-2.65)--(-5.7,-2.65) node[anchor=south east] {$\tau_{R}$};
\draw[->, gray, thick] (-6.2,-2.2) arc (-135:135:0.282843);
\draw[black, thick] (-3,-3)--(-4,-2)--(-2,-2)--cycle;
\filldraw[red] (-2,-2) circle (2pt);
\draw[red, very thick] (-3,-3)--(-2,-2);
\filldraw[blue] (-4,-2) circle (2pt);
\draw[blue, very thick] (-3,-3)--(-4,-2);
\draw[double, ->, >=stealth, black] (-2.7,-2.35)--(-3.3,-2.35) node[anchor=north west] {$\tau_{L}$};
\node[black] at (-2.3,-2.7) {$s_0$};
\node[black] at (-3.7,-2.7) {$s_1$};
\draw[->, gray, thick] (-2.2,-2.2) arc (225:-45:0.282843);
\draw[black, thick] (0,0)--(-1,1)--(-1,-1)--cycle;
\draw[black, thick] (0,0)--(-1,1)--(1,1)--cycle;
\draw[black, thick] (1,1)--(0,0)--(2,0)--cycle;
\draw[black, thick] (2,0)--(1,1)--(3,1)--cycle;
\draw[black, thick] (0,0)--(0,-2)--(-1,-1)--cycle;
\draw[black, thick] (0,-2)--(-1,-1)--(-1,-3)--cycle;
\draw[black, thick] (0,-2)--(-1,-3)--(1,-3)--cycle;
\draw[black, thick] (1,-3)--(0,-2)--(2,-2)--cycle;
\draw[black, thick] (2,-2)--(1,-3)--(3,-3)--cycle;
\draw[double, ->, >=stealth, black] (-0.65,-0.3)--(-0.65,0.3) node[anchor=north west] {\Romannum{1}};
\draw[double, ->, >=stealth, black] (0.7,0.35)--(1.3,0.35) node[anchor=south east] {\Romannum{2}};
\draw[double, ->, >=stealth, black] (-0.65,-1.7)--(-0.65,-2.3) node[anchor=south west] {\Romannum{3}};
\draw[double, ->, >=stealth, black] (0.7,-2.35)--(1.3,-2.35) node[anchor=north east] {\Romannum{4}};
\draw[red, very thick] (-1,-1)--(0,0);
\filldraw[red] (0,0) circle (2pt);
\node[black] at (-0.3,-0.7) {$s_0$};
\draw[blue, very thick] (2,0)--(3,1);
\filldraw[blue] (2,0) circle (2pt);
\node[black] at (2.7,0.3) {$s_1$};
\draw[blue, very thick] (2,-2)--(3,-3);
\filldraw[blue] (2,-2) circle (2pt);
\node[black] at (2.7,-2.3) {$s_1$};
\draw [double, ->, >=stealth, black] (0.2,-0.7) to [bend left] (1.7,-0.3);
\draw [double, ->, >=stealth, black] (0.2,-1) to [bend right] (1.7,-1.8);
\node[black] at (0.7,-0.7) {$\tau_{L}$};
\node[black] at (1.3,-1.3) {$\tau_{R}$};

\end{tikzpicture}
\end{tabular}

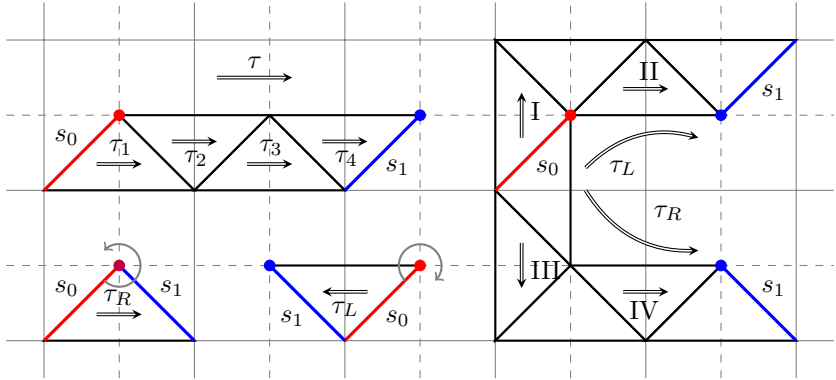
\captionof{figure}[foo]{A ribbon $\tau$ is composed of triangles $\tau_i$ $(i={1,2\cdots n})$ with a direction from $s_0$ to $s_1$. A triangle is a component of a ribbon with inherited direction and also the shortest ribbon.}
\label{tab:ribbon operators}
\end{table}

\subsection{Definition of ribbon operators}\label{subsec:definition of ribbon operators}
For a directed ribbon $\tau$ and $h \otimes f \in H \otimes H^*$, we will define the ribbon operator $F^{h \otimes f}(\tau)$, also written as $F^{(h,f)}(\tau)$. The operators will act on the whole Hilbert space $\mathcal{H}$, but the action is non-trivial only on the edges contained in $\tau$. Explicitly, for an elementary ribbon $\tau$, let $\mathcal{H}_{\tau}:= \mathcal{H}_{e_{\tau}}$ if $\tau$ is direct, and $\mathcal{H}_{\tau}:= \mathcal{H}_{e_{\tau}^*}$ otherwise. For a general ribbon $\tau$, decompose $\tau = \tau_1 \sqcup \tau_2$ so that $\partial_1 \tau_1 = \partial_0 \tau_2$ and define inductively $\mathcal{H}_{\tau}:= \mathcal{H}_{\tau_1} \otimes \mathcal{H}_{\tau_2}$. Then $F^{(h,f)}(\tau)$ will only act non-trivially on the space $\mathcal{H}_{\tau}$. The definition of ribbon operators below is motivated by \cite{kitaev2003fault} \cite{bombin2008family} for group algebras and by \cite{buerschaper2013hierarchy} for Hopf algebras. However, none of the above references addresses the critical issue of local orientation, as to be discussed later.

First, assume $\tau$ is an elementary directed ribbon, i.e., a triangle.
There are four cases depending on its type and local orientation. Also, recall that the edges in the lattice as well as those in the dual lattice are directed. The direction of the edge $e_{\tau}$  and that of $\tau$ can be either parallel or opposite. Taking this into consideration, we distinguish eight cases in Equations \ref{eqn:definitions of triangles}$a$-\ref{eqn:definitions of triangles}$h$, where Equations $(a)-(d)$ correspond to locally clockwise triangles and $(e)-(h)$  locally counterclockwise triangles. 

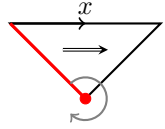
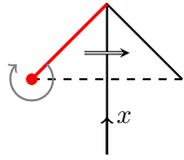
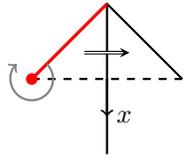
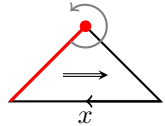
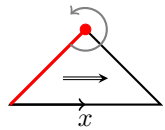
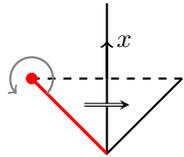
\begin{subequations}\label{eqn:definitions of triangles}
\begin{center}
\begin{tabular}{m{4cm} | m{7cm}}
\begin{center}
\begin{tikzpicture}
\draw[black, thick] (0,0)--(-1,1)--(1,1)--cycle;
\draw[->, black, thick] (1,1)--(0,1) node[anchor=south] {$x$};
\draw[double, ->, >=stealth, black] (-0.3,0.65)--(0.3,0.65);
\filldraw[red] (0,0) circle (2pt);
\draw[red, very thick] (-1,1)--(0,0);
\draw[->, gray, thick] (-0.2,0.2) arc (135:-135:0.282843);
\end{tikzpicture}
\end{center}
&
\begin{equation}
F^{(h,f)}(\tau_L)|x\rangle = \sum_{(x)} \epsilon(h)f[S(x'')]|x'\rangle
\end{equation}
\end{tabular}
\end{center}

\begin{center}
\begin{tabular}{m{4cm} | m{7cm}}
\begin{center}
\begin{tikzpicture}
\draw[black, thick] (0,0)--(-1,1)--(1,1)--cycle;
\draw[->, black, thick] (-1,1)--(0,1) node[anchor=south] {$x$};
\draw[double, ->, >=stealth, black] (-0.3,0.65)--(0.3,0.65);
\filldraw[red] (0,0) circle (2pt);
\draw[red, very thick] (-1,1)--(0,0);
\draw[->, gray, thick] (-0.2,0.2) arc (135:-135:0.282843);
\end{tikzpicture}
\end{center}
&
\begin{equation}
F^{(h,f)}(\tau_L)|x\rangle = \sum_{(x)} \epsilon(h)f(x')|x''\rangle
\end{equation}
\end{tabular}
\end{center}

\begin{center}
\begin{tabular}{m{4cm} | m{7cm}}
\begin{center}
\begin{tikzpicture}
\draw[black, thick] (0,0)--(1,1)--(2,0);
\draw[black, dashed, thick] (0,0)--(2,0);
\draw[black, thick] (1,-1)--(1,1);
\draw[->, black, thick] (1,-1)--(1,-0.5) node[anchor=west] {$x$};
\draw[double, ->, >=stealth, black] (0.7,0.35)--(1.3,0.35);
\filldraw[red] (0,0) circle (2pt);
\draw[red, very thick] (0,0)--(1,1);
\draw[->, gray, thick] (0.2,0.2) arc (45:-225:0.282843);
\end{tikzpicture}
\end{center}
&
\begin{equation}
F^{(h,f)}(\tau_L)|x\rangle = \epsilon(f)|xS(h)\rangle
\end{equation}
\end{tabular}
\end{center}

\begin{center}
\begin{tabular}{m{4cm} | m{7cm}}
\begin{center}
\begin{tikzpicture}
\draw[black, thick] (0,0)--(1,1)--(2,0);
\draw[black, dashed, thick] (0,0)--(2,0);
\draw[black, thick] (1,-1)--(1,1);
\draw[->, black, thick] (1,0)--(1,-0.5) node[anchor=west] {$x$};
\draw[double, ->, >=stealth, black] (0.7,0.35)--(1.3,0.35);
\filldraw[red] (0,0) circle (2pt);
\draw[red, very thick] (0,0)--(1,1);
\draw[->, gray, thick] (0.2,0.2) arc (45:-225:0.282843);
\end{tikzpicture}
\end{center}
&
\begin{equation}
F^{(h,f)}(\tau_L)|x\rangle = \epsilon(f)|hx\rangle
\end{equation}
\end{tabular}
\end{center}

\begin{center}
\begin{tabular}{m{4cm} | m{7cm}} 
\begin{center}
\begin{tikzpicture}
\draw[black, thick] (0,0)--(1,1)--(2,0)--cycle;
\draw[->, black, thick] (2,0)--(1,0) node[anchor=north] {$x$};
\draw[double, ->, >=stealth, black] (0.7,0.35)--(1.3,0.35);
\filldraw[red] (1,1) circle (2pt);
\draw[red, very thick] (0,0)--(1,1);
\draw[->, gray, thick] (0.8,0.8) arc (-135:135:0.282843);
\end{tikzpicture}
\end{center}
&
\begin{equation}
F^{(h,f)}(\tau_R)|x\rangle = \sum_{(x)} \epsilon(h)f[S(x')]|x''\rangle
\end{equation}
\end{tabular}
\end{center}

\begin{center}
\begin{tabular}{m{4cm} | m{7cm}}
\begin{center}
\begin{tikzpicture}
\draw[black, thick] (0,0)--(1,1)--(2,0)--cycle;
\draw[->, black, thick] (0,0)--(1,0) node[anchor=north] {$x$};
\draw[double, ->, >=stealth, black] (0.7,0.35)--(1.3,0.35);
\filldraw[red] (1,1) circle (2pt);
\draw[red, very thick] (0,0)--(1,1);
\draw[->, gray, thick] (0.8,0.8) arc (-135:135:0.282843);
\end{tikzpicture}
\end{center}
&
\begin{equation}
F^{(h,f)}(\tau_R)|x\rangle = \sum_{(x)} \epsilon(h)f(x'')|x'\rangle
\end{equation}
\end{tabular}
\end{center}

\begin{center}
\begin{tabular}{m{4cm} | m{7cm}}
\begin{center}
\begin{tikzpicture}
\draw[black, thick] (-1,1)--(0,0)--(1,1);
\draw[black, dashed, thick] (-1,1)--(1,1);
\draw[black, thick] (0,2)--(0,0);
\draw[->, black, thick] (0,1)--(0,1.5) node[anchor=west] {$x$};
\draw[double, ->, >=stealth, black] (-0.3,0.65)--(0.3,0.65);
\filldraw[red] (-1,1) circle (2pt);
\draw[red, very thick] (-1,1)--(0,0);
\draw[->, gray, thick] (-0.8,0.8) arc (-45:225:0.282843);
\end{tikzpicture}
\end{center}
&
\begin{equation}\label{eqn:g at s0}
F^{(h,f)}(\tau_R)|x\rangle = \epsilon(f)|S(h)x\rangle
\end{equation}
\end{tabular}
\end{center}

\begin{center}
\begin{tabular}{m{4cm} | m{7cm}}
\begin{center}
\begin{tikzpicture}
\draw[black, thick] (-1,1)--(0,0)--(1,1);
\draw[black, dashed, thick] (-1,1)--(1,1);
\draw[black, thick] (0,2)--(0,0);
\draw[->, black, thick] (0,2)--(0,1.5) node[anchor=west] {$x$};
\draw[double, ->, >=stealth, black] (-0.3,0.65)--(0.3,0.65);
\filldraw[red] (-1,1) circle (2pt);
\draw[red, very thick] (-1,1)--(0,0);
\draw[->, gray, thick] (-0.8,0.8) arc (-45:225:0.282843);
\end{tikzpicture}
\end{center}
&
\begin{equation}
F^{(h,f)}(\tau_R)|x\rangle = \epsilon(f)|xh\rangle
\end{equation}
\end{tabular}
\end{center}
\end{subequations}

For ribbons other than elementary triangles, we define the ribbon operators inductively. Let $\tau$ be an arbitrary ribbon. Decompose $\tau$ as $\tau=\tau_1 \sqcup \tau_2$, where the terminal site of $\tau_1$ matches the initial site of $\tau_2$, and they are disjoint otherwise. For $h \otimes f \in H \otimes H^*$, define
\begin{equation} \label{eqn:constrution of ribbon operator}
    F^{h,f}(\tau):=\sum_{i,(i),(h)} F^{h',g_i}(\tau_1) F^{S(i''')h''i',f(i''?)}(\tau_2),
\end{equation}
where $\{i\}$ is an orthogonal complete basis of $H$, and $g_i=\langle i,\;\rangle$ is the corresponding functional in $H^*$. The above definition is explicit, but a more intuitive way is as follows. For an element $h \otimes f \in D(H)^{*} \simeq H \otimes H^*$ where the isomorphism denotes a linear isomorphism between vector spaces, 
\begin{equation}
    \Delta(h \otimes f) =\sum_{(h \otimes f)} (h \otimes f)'\otimes (h \otimes f)''.
\end{equation}
We apply the expansion to the construction of ribbon operators as
\begin{equation}\label{eqn:constrution of ribbon operator2}
    F^{h \otimes f}(\tau):=\sum_{(h \otimes f)} F^{(h \otimes f)'}(\tau_1)F^{(h \otimes f)''}(\tau_2).
\end{equation}

It can be checked that Equations \ref{eqn:constrution of ribbon operator} and \ref{eqn:constrution of ribbon operator2} are equivalent. The ribbon operators do not depend on how the ribbon is partitioned into shorter ones due to the coassociativity of the comultiplication in Hopf algebras.

\subsection{Local orientation in original Kitaev model}\label{subsec:local orientation original}
In this subsection, we show that the distinction of local orientation is already necessary in the orignal Kitaev model. 
Note that, from Equations \ref{eqn:definitions of triangles},  $F^{(h,f)}(\tau)$ does not distinguish local orientations on direct triangles if $H$ is cocommutative, and it does not distinguish local orientations on dual triangles if $H$ is commutative. In particular, if $H$ is the group algebra of an Abelian group (e.g., toric code), then local orientations are redundant. On the other hand, for the group algebra of a non-Abelian group in the original Kitaev model, the two local orientations on a dual triangle should support different ribbon operators according to our definitions. This distinction, however,  has not been addressed in the literature, to the best of our knowledge. In \cite{kitaev2003fault}, \cite{bombin2008family}, the definition of ribbon operators on triangles  coincide with that presented in Equations \ref{eqn:definitions of triangles}$a$-\ref{eqn:definitions of triangles}$d$ corresponding to locally clockwise orientation. We show below with an explicit example that ignoring local orientations can cause certain properties to fail.

For the rest of the subsection, let $H = \mathbb{C}[G]$ be the group algebra of a non-Abelian group $G$. 
Equation \ref{eqn:F^h,g and B_th} is a commutation relation that is expected to hold between ribbon operators  and plaquette operators, where $s_0$ is the initial site of a ribbon $\tau$ (see Equation (B42) in \cite{bombin2008family}), and $t, h, g \in G$. 
\begin{equation}\label{eqn:F^h,g and B_th}
    B_t(s_0)F^{h,g}(\tau)=F^{h,g}(\tau)B_{th}(s_0).
\end{equation}
In fact, we just need the above identity to hold when both sides act on the ground state.

Take $\tau$ to be the ribbon shown in Figure \ref{tab:vialation}, which is a dual triangle and has locally counterclockwise orientation. In Appendix \ref{sec:Violation and correction in group algebra}, we show in detail that Equation \ref{eqn:F^h,g and B_th} fails for $\tau$ and any other ribbon that starts with $\tau$ if we use the old definition of ribbon operators on them.  By recognizing $\tau$ with locally counterclockwise orientation and using the new definition (Equation \ref{eqn:definitions of triangles}$g$), we can resolve the issue, and obtain the following commutation relation,  
\begin{equation}\label{eqn:F^h,g and B_ht}
    B_t(s_0)F^{h,g}(\tau)=F^{h,g}(\tau)B_{ht}(s_0),
\end{equation}
which is equivalent to Equation \ref{eqn:F^h,g and B_th} when acting on the ground state since $\delta_{ht,e}=\delta_{th,e}$.

\begin{table}[ht]
\begin{tabular}{m{11cm}}
\centering
\begin{tikzpicture}
\draw[gray, thick] (-1,-1) rectangle (1,1);
\draw[->, gray, thick] (0,-1)--(0.5,-1) node[anchor=north,black] {$x_1$};
\draw[->, gray, thick] (1,-1)--(1,0) node[anchor=west,black] {$x_2$};
\draw[->, gray, thick] (1,1)--(0,1) node[anchor=south,black] {$x_3$};
\draw[->, gray, thick] (-1,1)--(-1,0) node[anchor=east,black] {$x_4$};
\draw[black, thick] (0,0)--(-1,-1)--(0,-2);
\draw[black, dashed, thick] (0,0)--(0,-2);
\draw[double, ->, >=stealth, black] (-0.35,-0.7)--(-0.35,-1.3);
\filldraw[red] (0,0) circle (2pt);
\draw[red, very thick] (-1,-1)--(0,0);
\node[black] at (-0.7,-0.3) {$s_0$};
\end{tikzpicture}
\end{tabular}

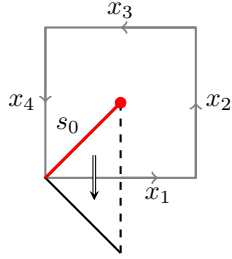
\captionof{figure}[foo]{A counter-example of a ribbon for which Equation \ref{eqn:F^h,g and B_th} fails in the original Kitaev model.}
\label{tab:vialation}
\end{table}

\subsection{Properties of ribbon operators}
\label{subsec:property_ribbon_operators}
In this section, we establish a few properties of ribbon operators. Recall that the ribbon operators $F^{h,f}(\tau)$ only act non-trivially on the Hilbert space $\mathcal{H}_{\tau}$. 
\begin{proposition}
Let $\tau_L$ and $\tau_R$ be a locally clockwise and a locally counterclockwise ribbon, respectively. Then, 
\begin{equation}
    F^{h_{1},f_{1}}(\tau_{L})\cdot F^{h_{2},f_{2}}(\tau_{L}) = F^{h_{1}h_{2},f_{2}f_{1}}(\tau_{L}),
\end{equation}
\begin{equation}\label{equ:multiplication_tauR}
    F^{h_{1},f_{1}}(\tau_{R})\cdot F^{h_{2},f_{2}}(\tau_{R}) = F^{h_{2}h_{1},f_{1}f_{2}}(\tau_{R}).
\end{equation}
In another words, the operators $F^{h,f}(\tau)$ define a representation of $D(H)^{*,op}$ on $\mathcal{H}_{\tau}$ if $\tau$ is locally clockwise, and a representation of $D(H)^*$ if $\tau$ is locally counterclockwise.
\begin{proof}
In Appendix \ref{sec:Representation of ribbons}, we show in details that the above two equations hold for elementary ribbons. Then it can be proved inductively that they also hold for general ribbons using the compatibility condition between multiplication and comultiplication in a Hopf algebra. Notice that $D(H)^{*,op}$ and $D(H)^*$ share the same comultiplication. Below we only give the proof for $\tau_R$ since that of the other case is similar.

Let $\tau_R$ be a locally counterclockwise ribbon. 
Assume Equation \ref{equ:multiplication_tauR} holds for any ribbon whose length is shorter than that of $\tau_R$. Decompose $\tau_R$ as $\tau_R = \tau_1 \sqcup \tau_2$ such that $\partial_1 \tau_1 = \partial_0 \tau_2$. Then,
\begin{align*}
&{}\quad\  F^{h_1 \otimes f_1}(\tau_R) \cdot F^{h_2 \otimes f_2}(\tau_R) \\
&= \sum_{(h_1 \otimes f_1)} F^{(h_1 \otimes f_1)'}(\tau_1)F^{(h_1 \otimes f_1)''}(\tau_2) \ \cdot \ \sum_{(h_2 \otimes f_2)} F^{(h_2 \otimes f_2)'}(\tau_1)F^{(h_2 \otimes f_2)''}(\tau_2) \\
&= \sum_{(h_1 \otimes f_1)}\sum_{(h_2 \otimes f_2)} F^{(h_1 \otimes f_1)'}(\tau_1)F^{(h_2 \otimes f_2)'}(\tau_1) \ F^{(h_1 \otimes f_1)''}(\tau_2)F^{(h_2 \otimes f_2)''}(\tau_2) \\
&= \sum_{(h_1 \otimes f_1)}\sum_{(h_2 \otimes f_2)} F^{(h_1 \otimes f_1)'(h_2 \otimes f_2)'}(\tau_1)\ F^{(h_1 \otimes f_1)''(h_2 \otimes f_2)''}(\tau_2) \\
&= \sum_{(h_2h_1 \otimes f_1f_2)}F^{(h_2h_1 \otimes f_1f_2)'}(\tau_1)\ F^{(h_2h_1 \otimes f_1f_2)''}(\tau_2) \\
&= F^{h_{2}h_{1},f_{1}f_{2}}(\tau_{R}).
\end{align*}
In the above derivation, the first and the last equality are due to Equation \ref{eqn:constrution of ribbon operator2}, the third by induction, the fourth by the compatibility condition between multiplication and comultiplication in $D(H)^*$, and the second by the commutativity between ribbon operators on $\tau_1$ and those on $\tau_2$.
\end{proof}
\end{proposition}

Next, we examine the commutation relation between ribbon operators and local operators. Let $\ket{GS} \in \mathcal{H}$ be the ground state\footnote{To the interest of the current paper, we can assume the lattice is defined on the sphere or the infinite plane, and so there is a unique ground state.}. Then at any site $s$, the local operators act on $\ket{GS}$ as follows,
\begin{align*}
A_a(s) \ket{GS} &= \ket{GS},\\
B_f(s) \ket{GS} &= f(1)\ket{GS}, \quad a \in H, f \in H^*. 
\end{align*}
Let $\tau$ be a ribbon with initial site $s_0 = \partial_0 \tau$ and terminal site $s_1 = \partial_1 \tau$. Assume the length of $\tau$, i.e., the number of triangles contained in $\tau$, is greater than one. The following is a technical lemma concerning the commutation relation between ribbon operators on $\tau$ and local operators on its ends.
\begin{lemma}\label{lem:local_operator_at_ends}
Let $\tau_L$ and $\tau_R$ be a locally clockwise and a locally counterclockwise ribbon, respectively, as described above. 
\begin{enumerate}[label = {(\arabic*)}]
    \item At $s_0$, we have
    \begin{subequations}\label{eqn:Calculation Relation at s0}
\begin{align}
& A_{a}(s_{0})F^{(h,f)}(\tau_{L}) = \sum_{(a)} F^{\{a'hS(a'''),f[S(a'')?]\}}(\tau_{L}) A_{a^{(4)}}(s_0), \label{eqn:a at s0}\\
& A_{a}(s_{0})F^{(h,f)}(\tau_{R}) = \sum_{(a)} F^{\{a''hS(a^{(4)}),f[S(a''')?]\}}(\tau_{R}) A_{a'}(s_{0}), \label{eqn:b at s0}\\
& B_{t}(s_{0})F^{(h,f)}(\tau_{L}) = \sum_{(h)} F^{(h'',f)}(\tau_{L}) B_{t[?S(h')]}(s_{0}), \label{eqn:c at s0}\\
& B_{t}(s_{0})F^{(h,f)}(\tau_{R}) = \sum_{(h)} F^{(h'',f)}(\tau_{R}) B_{t[S(h')?]}(s_{0}). \label{eqn:d at s0}
\end{align}
\end{subequations}
\item At $s_1$, we have
\begin{subequations}\label{eqn:Calculation Relation at s1}
\begin{align}
& A_{a}(s_{1})F^{(h,f)}(\tau_{L}) = \sum_{(a)} F^{[h,f(?a'')]}(\tau_{L}) A_{a'}(s_{1}), \label{eqn:a at s1}\\
& A_{a}(s_{1})F^{(h,f)}(\tau_{R}) = \sum_{(a)} F^{[h,f(?a')]}(\tau_{R}) A_{a''}(s_{1}), \label{eqn:b at s1}\\
& B_{t}(s_{1})F^{(h,f)}(\tau_{L}) = \sum_{(i),(h),i} f(i'') F^{(h',g_{i})}(\tau_{L}) B_{t[S(i''')h''i'?]}(s_{1}), \label{eqn:c at s1}\\
& B_{t}(s_{1})F^{(h,f)}(\tau_{R}) = \sum_{(i),(h),i} f(i'') F^{(h',g_{i})}(\tau_{R}) B_{t[?S(i''')h''i']}(s_{1}). \label{eqn:d at s1}
\end{align}
\end{subequations}
In the above, $\{i\}$ is an orthogonal complete basis of $H$, and $g_i=\langle i,\;\rangle$ is the corresponding functional in $H^*$.
\end{enumerate}
\begin{proof}
For a detailed proof, see Appendix \ref{sec:Local actions on ribbon operators}. The idea is that we first prove the above equations for ribbons with shortest possible length, and then extend the equality to longer ribbons using the decomposition formula in Equation \ref{eqn:constrution of ribbon operator2}.
The shortest possible ribbons for the equalities in Equation \ref{eqn:Calculation Relation at s0} are illustrated in Figure \ref{tab:Commutation Relation at s0}, and those in Equation \ref{eqn:Calculation Relation at s1} illustrated in Figure \ref{tab:Commutation Relation at s1}.
\end{proof}
\end{lemma}
\begin{table}[ht]
\centering
\begin{tabular}{m{5.5cm} | m{5.5cm}} 
\begin{center}
\begin{tikzpicture}
\draw[gray, thick] (0,-1.2)--(0,1.6);
\draw[gray, thick] (-2,0)--(2,0);
\draw[->, gray, thick] (0,0)--(1,0) node[anchor=north,black] {$x_1$};
\draw[->, gray, thick] (0,-1)--(0,-0.8) node[anchor=west,black] {$x_2$};
\draw[->, gray, thick] (0,0)--(-1,0) node[anchor=north,black] {$x_3$};
\draw[->, gray, thick] (0,1.6)--(0,1.3) node[anchor=east,black] {$x_4$};
\draw[black, thick] (0,0)--(-1,1)--(-2,0)--cycle;
\draw[black, thick] (0,0)--(1,1)--(2,0)--cycle;
\draw[black, dashed, thick] (1,1)--(-1,1);
\draw[double, ->, >=stealth, black] (-0.2,0.5)--(-0.8,0.5) node[anchor=north west] {$(a)$};
\draw[double, ->, >=stealth, black] (0.7,0.35)--(1.3,0.35) node[anchor=south east] {$(b)$};
\filldraw[red] (1,1) circle (2pt);
\node[black] at (0.3,0.7) {$s_0$};
\draw[red, very thick] (0,0)--(1,1);
\end{tikzpicture}
\end{center}
&
\begin{center}
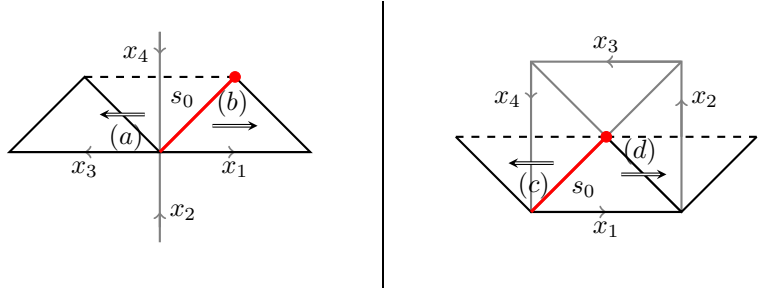

\begin{tikzpicture}
\draw[gray, thick] (-1,-1) rectangle (1,1);
\draw[gray, thick] (-1,1)--(1,-1);
\draw[gray, thick] (-1,-1)--(1,1);
\draw[->, gray, thick] (-1,-1)--(0,-1) node[anchor=north,black] {$x_1$};
\draw[->, gray, thick] (1,0)--(1,0.5) node[anchor=west,black] {$x_2$};
\draw[->, gray, thick] (1,1)--(0,1) node[anchor=south,black] {$x_3$};
\draw[->, gray, thick] (-1,1)--(-1,0.5) node[anchor=east,black] {$x_4$};
\draw[black, thick] (0,0)--(-1,-1)--(-2,0);
\draw[black, dashed, thick] (0,0)--(-2,0);
\draw[black, thick] (0,0)--(1,-1)--(2,0);
\draw[black, dashed, thick] (0,0)--(2,0);
\draw[black, thick] (-1,-1)--(1,-1);
\draw[double, ->, >=stealth, black] (-0.7,-0.35)--(-1.3,-0.35) node[anchor=north west] {$(c)$};
\draw[double, ->, >=stealth, black] (0.2,-0.5)--(0.8,-0.5) node[anchor=south east] {$(d)$};
\filldraw[red] (0,0) circle (2pt);
\node[black] at (-0.3,-0.7) {$s_0$};
\draw[red, very thick] (-1,-1)--(0,0);
\end{tikzpicture}
\end{center}
\end{tabular}
\captionof{figure}[foo]{Ribbons marked with (a)-(d) correspond the Equation \ref{eqn:Calculation Relation at s0} a-d.} 
\label{tab:Commutation Relation at s0}
\end{table}
\begin{table}[ht]
\centering
\begin{tabular}{m{5.5cm} | m{5.5cm}} 
\begin{center}
\begin{tikzpicture}
\draw[gray, thick] (0,-1.2)--(0,1.6);
\draw[gray, thick] (-2,0)--(2,0);
\draw[->, gray, thick] (0,0)--(1,0) node[anchor=north,black] {$x_1$};
\draw[->, gray, thick] (0,-1)--(0,-0.8) node[anchor=west,black] {$x_2$};
\draw[->, gray, thick] (0,0)--(-1,0) node[anchor=north,black] {$x_3$};
\draw[->, gray, thick] (0,1.6)--(0,1.3) node[anchor=east,black] {$x_4$};
\draw[black, thick] (0,0)--(-1,1)--(-2,0)--cycle;
\draw[black, thick] (0,0)--(1,1)--(2,0)--cycle;
\draw[black, dashed, thick] (1,1)--(-1,1);
\draw[double, ->, >=stealth, black] (1.3,0.35)--(0.7,0.35) node[anchor=south west] {$(a)$};
\draw[double, ->, >=stealth, black] (-0.8,0.5)--(-0.2,0.5) node[anchor=north east] {$(b)$};
\filldraw[blue] (1,1) circle (2pt);
\node[black] at (0.3,0.7) {$s_1$};
\draw[blue, very thick] (0,0)--(1,1);
\end{tikzpicture}
\end{center}
&
\begin{center}
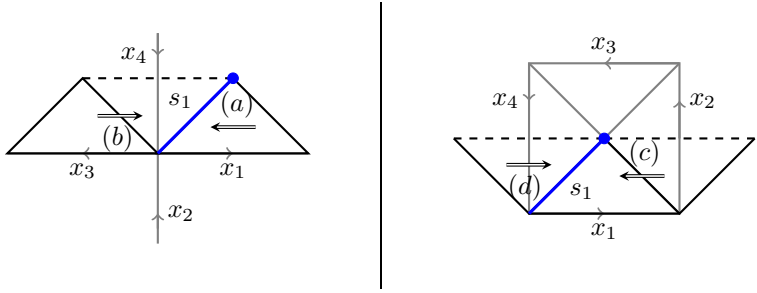

\begin{tikzpicture}
\draw[gray, thick] (-1,-1) rectangle (1,1);
\draw[gray, thick] (-1,1)--(1,-1);
\draw[gray, thick] (-1,-1)--(1,1);
\draw[->, gray, thick] (-1,-1)--(0,-1) node[anchor=north,black] {$x_1$};
\draw[->, gray, thick] (1,0)--(1,0.5) node[anchor=west,black] {$x_2$};
\draw[->, gray, thick] (1,1)--(0,1) node[anchor=south,black] {$x_3$};
\draw[->, gray, thick] (-1,1)--(-1,0.5) node[anchor=east,black] {$x_4$};
\draw[black, thick] (0,0)--(-1,-1)--(-2,0);
\draw[black, dashed, thick] (0,0)--(-2,0);
\draw[black, thick] (0,0)--(1,-1)--(2,0);
\draw[black, dashed, thick] (0,0)--(2,0);
\draw[black, thick] (-1,-1)--(1,-1);
\draw[double, ->, >=stealth, black] (0.8,-0.5)--(0.2,-0.5) node[anchor=south west] {$(c)$};
\draw[double, ->, >=stealth, black] (-1.3,-0.35)--(-0.7,-0.35) node[anchor=north east] {$(d)$};
\filldraw[blue] (0,0) circle (2pt);
\node[black] at (-0.3,-0.7) {$s_1$};
\draw[blue, very thick] (-1,-1)--(0,0);
\end{tikzpicture}
\end{center}
\end{tabular}
\captionof{figure}[foo]{Ribbons marked with (a)-(d) correspond the Equation \ref{eqn:Calculation Relation at s1} a-d.}
\label{tab:Commutation Relation at s1}
\end{table}

Using Lemma \ref{lem:local_operator_at_ends}, we can also deduce that ribbon operators commute with all terms in the Hamiltonian except for those associated with the ends of the ribbon.
\begin{proposition}\label{prop:ribbon_operator_in_middle}
Let $\tau$ be a ribbon and $s$ be a site on $\tau$ such that $s$ has no overlap with $\partial_i \tau$. Denote the terms associated to $s$ in the Hamiltonian by $A(s)=A_{h_0}(s) , B(s)=B_{\phi}(s)$ where $h_0 \in H$ is the Haar integral of $H$ and $\phi \in H^*$ is the Haar integral of $H^*$. Then,
\begin{subequations}\label{eqn:Calculation Relation at middle}
\begin{align}
A(s) F^{(h,f)}(\tau) &= F^{(h,f)}(\tau) A(s), \label{eqn:a at middle}\\
B(s) F^{(h,f)}(\tau) &= F^{(h,f)}(\tau) B(s). \label{eqn:b at middle}
\end{align}
\end{subequations}
\begin{proof}
See Appendix \ref{sec:Calculation on commutation relation with ribbon in middle} for a proof.
\end{proof}
\end{proposition}

The commutation relation between ribbon operators and local operators at the ends in Lemma \ref{lem:local_operator_at_ends} may look complicated. However, if we restrict ribbon operators on the ground state, then those relations reduce to more compact formulas. Let $V_{\tau}$ be the Hilbert space of ribbon operators on  $\tau$ acting on the ground state, 
\begin{align*}
V_{\tau} = \text{span}_{\mathbb{C}}\{\ket{h \otimes f} \equiv F^{h \otimes f}(\tau)\ket{GS}\ :\ h \otimes f \in D(H)^*\}.   
\end{align*}
Then, $V_{\tau}$ is naturally identified with the space $D(H)^*$. Recall from Equations \ref{eqn:left action on H^*} and \ref{eqn:right action on H^*}, $D(H)$, as a Hopf algebra, has two natural representations on $D(H)^*$ denoted by $L$ and $R$, where $L$ is induced from the left multiplication of $D(H)$ on itself and $R$ is induced from the right multiplication (precomposed by the antipode). Apparently, these two actions commute with each other.
\begin{proposition}
Let $\tau$ be a ribbon of either local orientation with $s_i = \partial_i \tau$. Identify $V_{\tau}$ with $D(H)^*$. Then the local operators $B_t(s_0)A_a(s_0)$ define a representation of $D(H)$ on $V_{\tau}$ isomorphic to $L$, and $B_t(s_1)A_a(s_1)$ define a representation isomorphic to $R$.
\begin{proof}
 The statement can be proved by restricting the identities in Equations \ref{eqn:Calculation Relation at s0} and \ref{eqn:Calculation Relation at s1} on the ground state. It is straightforward to see that, at $s_0$, the two identities in Equations \ref{eqn:a at s0} and \ref{eqn:b at s0} corresponding to the two cases of local orientations both reduce to,
 \begin{equation} \label{eqn:behavior on ground state}
    A_a(s_0)|h\otimes f\rangle = \sum_{(a)} |a'hS(a'''),f\left[S(a'')?\right]\rangle,
\end{equation}
which agrees with Equation \ref{eqn:left action on H^*}, the action $L$ on $D(H)^*$:
\begin{equation}
    L(a)|(h\otimes f)\rangle=|(h \otimes f)(S(a)?)\rangle=\sum_{(a)}|a'hS(a''')\otimes f\left[S(a'')?\right] \rangle.
\end{equation}
Similarly, at $s_1$, for either local orientation we have
\begin{equation}
    A_a(s_1)|h\otimes f\rangle=|h,f(?a)\rangle,
\end{equation}
which agrees with Equation \ref{eqn:right action on H^*}, the action $R$ on $D(H)^*$:
\begin{equation}
    R(a)|h\otimes f\rangle=|(h \otimes f)(?a)\rangle= |h \otimes f(?a)\rangle.
\end{equation}
We leave the verification for the actions of $B_f(s_0)$ and $B_f(s_1)$ as an exercise.
\end{proof}
\end{proposition}

To summarize, ribbon operators on a sufficiently long ribbon $\tau$ commute with all terms in the Hamiltonian except those associated with the ends of $\tau$. Hence, ribbon operators create excitations only at the ends of a ribbon. When acting on the ground state, the space of ribbon operators on $\tau$ is naturally identified with $D(H)^*$. The action of local operators on $\partial_i \tau$ preserve $D(H)^*$. Thus, $D(H)^*$ can be thought of as the space of elementary excitations. More specifically, the action on $\partial_0 \tau$ define a representation of $D(H)$ on $D(H)^*$ coinciding with $L$, and that on $\partial_1 \tau$ a representation of  $D(H)$ on $D(H)^*$ coinciding with $R$. These two actions commute. By standard representation theory (see Equation \ref{equ:H_regular_decomposition}), we have the decomposition,
\begin{equation}
    D(H)^* \simeq \bigoplus_{\mu \in \text{Irr}_{D(H)}} \mu \otimes \mu^*,
\end{equation}
where $L$ acts on the first factor and $R$ acts on the second factor. Therefore, the local operators on the ends of $\tau$ can map a state in a sector $\mu^* \otimes \mu$ to any other state within the same sector, but cannot permute states of different sectors. This implies that the types of elementary excitations are labelled by irreducible representations of $D(H)$. Using Fourier transformation, it is not hard to find a specific basis $\{\langle \nu ab|\ :\ \nu \in \text{Irr}_{D(H)}, a,b = 1, \cdots, dim(\nu)\}$ of $D(H)^*$ so that $L$ acts only on the $a$ index and $R$ acts only on the $b$ index (See Appendix \ref{sec:FT of $H^*$}). That is, for $m \in D(H)$,
\begin{align}
L(m)(\langle \nu a b|) &= \sum_{k}D^{\nu} (m)_{ka} \langle\nu k b|, \label{equ:dual_basis_L_act}\\
R(m)(\langle \nu ab|) &= \sum_{k} D^{\nu^*}(m)_{kb}\langle \nu a k|. \label{equ:dual_basis_R_act}
\end{align}

\section{Conclusion and outlook}
In this paper, we provided a concrete definition of ribbon operators in the generalized Kitaev quantum double model, which is constructed over a semisimple Hopf algebra. We introduced the notion of local orientation on ribbons which we must distinguish in defining the operators on them. It was shown that even in the original Kitaev model based on non-Abelian groups, the issue of local orientation has to be addressed. Otherwise, certain properties of ribbon operators that are expected to hold would fail. We derived some properties of ribbon operators in the generalized model. For instance, they create quasi-particle excitations only at the end of the ribbon, and the types of the quasi-particles correspond to irreducible representations of the Drinfeld double of the input Hopf algebra. While these properties are a folklore, their derivations are technically complicated.

There are several future directions to proceed. Firstly, since this Hopf-algebra-model can be further replaced by a weak Hopf algebra (or quantum groupoid) \cite{chang2014kitaev}, it will be interesting to define and study ribbon operators in that case. Secondly, the generalized Kitaev model may find applications in topological quantum computing. For example, which Hopf algebras support universal quantum computing? Lastly, in \cite{cong2017hamiltonian}, the authors gave a Hamiltonian formulation for gapped boundaries in the original Kitaev model.  It will be interesting to generalize the formulation to the case of Hopf algebras.

\vspace{0.5cm}
\noindent \textbf{Acknowledgments.} 
The authors are partially supported by NSF CCF 2006667 and ARO MURI.  

\vspace{0.5cm}
\noindent \textbf{Data Availability.} 
The data that supports the findings of this study are available within the article.

\bibliographystyle{plain}
\bibliography{Kitaev_bib}

\appendix
\section{Straightening equation of $A_a$ and $B_f$}\label{sec:Straightening equation of $A_a$ and $B_f$}
\begin{table}[ht]
\centering
\begin{tabular}{m{4.3cm} | m{3.2cm} | m{3.2cm}}
\begin{tikzpicture}
\draw[black, thick] (-1,-1) rectangle (1,1);
\draw[black, thick] (-1,-1)--(-2.5,-1);
\draw[black, thick] (-1,-1)--(-1,-2.5);
\draw[->, black, thick] (-1,-1)--(0,-1) node[anchor=north] {$x_1$};
\draw[->, black, thick] (1,-1)--(1,0) node[anchor=west] {$x_2$};
\draw[->, black, thick] (1,1)--(0,1) node[anchor=south] {$x_3$};
\draw[->, black, thick] (-1,1)--(-1,0) node[anchor=east] {$x_4$};
\draw[->, black, thick] (-1,-1)--(-2,-1) node[anchor=north] {$x_5$};
\draw[->, black, thick] (-1,-2.5)--(-1,-2) node[anchor=east] {$x_6$};
\filldraw[red] (0,0) circle (2pt);
\node[black] at (-0.3,-0.7) {$s$};
\draw[red, very thick] (-1,-1)--(0,0);
\end{tikzpicture}
&
\begin{tikzpicture} 
\draw[black, thick] (-1.5,0)--(1.5,0);
\draw[black, thick] (0,-1.5)--(0,1.5);
\draw[->, black, thick] (0,0)--(0,1) node[anchor=south east] {$x_4$};
\draw[->, black, thick] (0,0)--(-1,0) node[anchor=north east] {$x_5$};
\draw[->, black, thick] (0,0)--(0,-1) node[anchor=north west] {$x_6$};
\draw[->, black, thick] (0,0)--(1,0) node[anchor=south west] {$x_1$};
\filldraw[red] (1,1) circle (2pt);
\draw[red, very thick] (0,0)--(1,1);
\node[black] at (0.3,0.7) {$s$};
\draw[->, black, thick] (0.12941,0.48296) arc (75:375:0.5);
\end{tikzpicture}
&
\begin{tikzpicture} 
\draw[black, thick] (-1,-1) rectangle (1,1);
\draw[->, black, thick] (-0.5,-1)--(0,-1) node[anchor=north] {$x_1$};
\draw[->, black, thick] (1,-1)--(1,0) node[anchor=west] {$x_2$};
\draw[->, black, thick] (1,1)--(0,1) node[anchor=south] {$x_3$};
\draw[->, black, thick] (-1,1)--(-1,0) node[anchor=east] {$x_4$};
\filldraw[red] (0,0) circle (2pt);
\draw[red, very thick] (-1,-1)--(0,0);
\node[black] at (-0.7,-0.3) {$s$};
\draw[->, black, thick] (-0.12941,-0.48296) arc (-105:195:0.5);
\end{tikzpicture}
\end{tabular}
\end{table}

This equation holds no matter how the edges are oriented. We check the case as shown  above.
\begin{align*}
&\quad A_{a}(s) B_{f}(s) |x_{1} \; x_{2} \; x_{3} \; x_{4} \; x_{5} \; x_{6}\rangle\\
&= A_{a}(s) \sum_{(x_i)} f(x_{1}'' x_{2}'' x_{3}'' x_{4}'') |x_{1}' \; x_{2}' \; x_{3}' \; x_{4}' \; x_{5} \; x_{6}\rangle\\
&= \sum_{(x_i),(a)} f(x_{1}'' x_{2}'' x_{3}'' x_{4}'') |a^{(4)}x_{1}' \; x_{2}' \; x_{3}' \; x_{4}'S(a') \; a''x_{5} \; x_{6}S(a''')\rangle\\
&= \sum_{(x_i),(a)} f[S(a^{(8)})a^{(7)}x_{1}'' x_{2}'' x_{3}'' x_{4}''S(a'')a'] &\\
&\qquad \qquad |a^{(6)}x_{1}' \; x_{2}' \; x_{3}' \; x_{4}'S(a''') \; a^{(4)}x_{5}' \; x_{6}S(a^{(5)})\rangle\\
&= \sum_{(x_i),(a)} B_{f[S(a^{(6)})?a']}(s) |a^{(5)}x_{1}\; x_{2} \; x_{3} \; x_{4}S(a'') \; a'''x_{5} \; x_{6}S(a^{(4)})\rangle\\
&= \sum_{(a)} B_{f[S(a''')?a']}(s) A_{a''}(s) |x_{1} \; x_{2} \; x_{3} \; x_{4} \; x_{5} \; x_{6}\rangle
\end{align*}
This is exactly the straightening equation. 

\section{Violation and correction in group algebra}\label{sec:Violation and correction in group algebra}
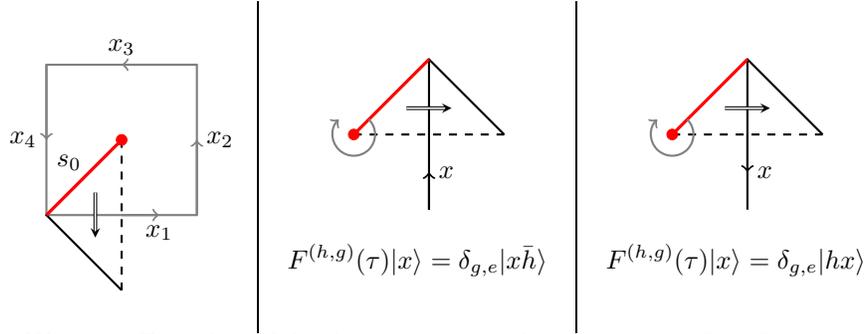
\begin{tabular}{m{3.2cm} | m{3.8cm} | m{3.8cm}} 
\begin{center}
\begin{tikzpicture}
\draw[gray, thick] (-1,-1) rectangle (1,1);
\draw[->, gray, thick] (0,-1)--(0.5,-1) node[anchor=north,black] {$x_1$};
\draw[->, gray, thick] (1,-1)--(1,0) node[anchor=west,black] {$x_2$};
\draw[->, gray, thick] (1,1)--(0,1) node[anchor=south,black] {$x_3$};
\draw[->, gray, thick] (-1,1)--(-1,0) node[anchor=east,black] {$x_4$};
\draw[black, thick] (0,0)--(-1,-1)--(0,-2);
\draw[black, dashed, thick] (0,0)--(0,-2);
\draw[double, ->, >=stealth, black] (-0.35,-0.7)--(-0.35,-1.3);
\filldraw[red] (0,0) circle (2pt);
\node[black] at (-0.7,-0.3) {$s_0$};
\draw[red, very thick] (-1,-1)--(0,0);
\end{tikzpicture}
\end{center}
&
\begin{center}
\begin{tikzpicture}
\draw[black, thick] (0,0)--(1,1)--(2,0);
\draw[black, dashed, thick] (0,0)--(2,0);
\draw[black, thick] (1,-1)--(1,1);
\draw[->, black, thick] (1,-1)--(1,-0.5) node[anchor=west] {$x$};
\draw[double, ->, >=stealth, black] (0.7,0.35)--(1.3,0.35);
\filldraw[red] (0,0) circle (2pt);
\draw[red, very thick] (0,0)--(1,1);
\draw[->, gray, thick] (0.2,0.2) arc (45:-225:0.282843);
\end{tikzpicture}
\begin{equation*}
F^{(h,g)}(\tau)|x\rangle = \delta_{g,e}|x\bar{h}\rangle
\end{equation*}
\end{center}
&
\begin{center}
\begin{tikzpicture}
\draw[black, thick] (0,0)--(1,1)--(2,0);
\draw[black, dashed, thick] (0,0)--(2,0);
\draw[black, thick] (1,-1)--(1,1);
\draw[->, black, thick] (1,0)--(1,-0.5) node[anchor=west] {$x$};
\draw[double, ->, >=stealth, black] (0.7,0.35)--(1.3,0.35);
\filldraw[red] (0,0) circle (2pt);
\draw[red, very thick] (0,0)--(1,1);
\draw[->, gray, thick] (0.2,0.2) arc (45:-225:0.282843);
\end{tikzpicture}
\begin{equation*}
F^{(h,g)}(\tau)|x\rangle = \delta_{g,e}|hx\rangle
\end{equation*}
\end{center}
\end{tabular}

We show Equation \ref{eqn:F^h,g and B_th} is violated for the ribbon $\tau$ in the first figure above for the original Kitaev model where $H$ is taken to be the group algebra of a non-Abelian group $G$. In \cite{bombin2008family}, only two formulas are provided for dual triangles as shown in the second and third figure above.  However, we can not get the desired commutation relation using either of them: 

\begin{align*}
&\quad B_{h'}(s_0) F^{(h,g)}(\tau) |x_{1} \; x_{2} \; x_{3} \; x_{4}\rangle\\
&= B_{h'}(s_0) \delta_{g,e} |x_{1}\bar{h} \; x_{2} \; x_{3} \; x_{4}\rangle\\
&= \delta_{h',x_{1}\bar{h}x_{2}x_{3}x_{4}} \delta_{g,e} |x_{1}\bar{h} \; x_{2} \; x_{3} \; x_{4}\rangle\\
&\neq \delta_{g,e} \delta_{hh',x_{1}x_{2}x_{3}x_{4}} |x_{1}\bar{h} \; x_{2} \; x_{3} \; x_{4}\rangle\\
&= F^{(h,g)}(\tau) \delta_{hh',x_{1}x_{2}x_{3}x_{4}} |x_{1} \; x_{2} \; x_{3} \; x_{4}\rangle\\
&= F^{(h,g)}(\tau) B_{hh'}(s_0) |x_{1} \; x_{2} \; x_{3} \; x_{4}\rangle\\
\\
&\quad B_{h'}(s_0) F^{(h,g)}(\tau) |x_{1} \; x_{2} \; x_{3} \; x_{4}\rangle\\
&= B_{h'}(s_0) \delta_{g,e} |hx_{1} \; x_{2} \; x_{3} \; x_{4}\rangle\\
&= \delta_{h',hx_{1}x_{2}x_{3}x_{4}} \delta_{g,e} |hx_{1} \; x_{2} \; x_{3} \; x_{4}\rangle\\
&\neq \delta_{g,e} \delta_{hh',x_{1}x_{2}x_{3}x_{4}} |hx_{1} \; x_{2} \; x_{3} \; x_{4}\rangle\\
&= F^{(h,g)}(\tau) \delta_{hh',x_{1}x_{2}x_{3}x_{4}} |x_{1} \; x_{2} \; x_{3} \; x_{4}\rangle\\
&= F^{(h,g)}(\tau) B_{hh'}(s_0) |x_{1} \; x_{2} \; x_{3} \; x_{4}\rangle
\end{align*}

Moreover, the issue can not be removed by making $\tau$ longer. Roughly, this is because for the current $\tau$, the initial site and terminal site already lie in different plaquettes, and thus lengthening it will not affect the action of the plaquette operator at the initial site.

To resolve the issue, we recognize that $\tau$ has locally counterclockwise orientation, and hence we need to apply the following formulas for the ribbon operators,
\begin{tabular}{m{5cm} | m{5cm}} 
\begin{center}
\begin{tikzpicture}
\draw[black, thick] (-1,1)--(0,0)--(1,1);
\draw[black, dashed, thick] (-1,1)--(1,1);
\draw[black, thick] (0,2)--(0,0);
\draw[->, black, thick] (0,1)--(0,1.5) node[anchor=west] {$x$};
\draw[double, ->, >=stealth, black] (-0.3,0.65)--(0.3,0.65);
\filldraw[red] (-1,1) circle (2pt);
\draw[red, very thick] (-1,1)--(0,0);
\draw[->, gray, thick] (-0.8,0.8) arc (-45:225:0.282843);
\end{tikzpicture}
\begin{equation*}
F^{(h,g)}(\tau)|x\rangle = \delta_{g,e}|\bar{h}x\rangle
\end{equation*}
\end{center}
&
\begin{center}
\begin{tikzpicture}
\draw[black, thick] (-1,1)--(0,0)--(1,1);
\draw[black, dashed, thick] (-1,1)--(1,1);
\draw[black, thick] (0,2)--(0,0);
\draw[->, black, thick] (0,2)--(0,1.5) node[anchor=west] {$x$};
\draw[double, ->, >=stealth, black] (-0.3,0.65)--(0.3,0.65);
\filldraw[red] (-1,1) circle (2pt);
\draw[red, very thick] (-1,1)--(0,0);
\draw[->, gray, thick] (-0.8,0.8) arc (-45:225:0.282843);
\end{tikzpicture}
\begin{equation*}
F^{(h,g)}(\tau)|x\rangle = \delta_{g,e}|xh\rangle
\end{equation*}
\end{center}
\end{tabular}

With the new formula above, we have,
\begin{eqnarray*}
&\quad& B_{h'}(s_0) F^{(h,g)}(\tau) |x_{1} \; x_{2} \; x_{3} \; x_{4}\rangle\\
&=& B_{h'}(s_0) \delta_{g,e} |\bar{h}x_{1} \; x_{2} \; x_{3} \; x_{4}\rangle\\
&=& \delta_{h',\bar{h}x_{1}x_{2}x_{3}x_{4}} \delta_{g,e} |\bar{h}x_{1} \; x_{2} \; x_{3} \; x_{4}\rangle\\
&=& \delta_{g,e} \delta_{hh',x_{1}x_{2}x_{3}x_{4}} |\bar{h}x_{1} \; x_{2} \; x_{3} \; x_{4}\rangle\\
&=& F^{(h,g)}(\tau) \delta_{hh',x_{1}x_{2}x_{3}x_{4}} |x_{1} \; x_{2} \; x_{3} \; x_{4}\rangle\\
&=& F^{(h,g)}(\tau) B_{hh'}(s_0) |x_{1} \; x_{2} \; x_{3} \; x_{4}\rangle
\end{eqnarray*}

\section{Multiplication of ribbon operators on elementary ribbons}\label{sec:Representation of ribbons}
\subsection{For locally clockwise ribbons $\tau_L$}
\begin{center}
\begin{tabular}{m{4cm} | m{7cm}}
\begin{center}
\begin{tikzpicture}
\draw[black, thick] (0,0)--(-1,1)--(1,1)--cycle;
\draw[->, black, thick] (1,1)--(0,1) node[anchor=south] {$x$};
\draw[double, ->, >=stealth, black] (-0.3,0.65)--(0.3,0.65);
\filldraw[red] (0,0) circle (2pt);
\draw[red, very thick] (-1,1)--(0,0);
\draw[->, gray, thick] (-0.2,0.2) arc (135:-135:0.282843);
\end{tikzpicture}
\end{center}
&
\begin{eqnarray*}
&\;& F^{(h_{1},f_{1})}(\tau_L) F^{(h_{2},f_{2})}(\tau_L) |x\rangle\\ 
&=& \sum_{(x)} F^{(h_{1},f_{1})}(\tau_L) \epsilon(h_{2})f_{2}[S(x'')]|x'\rangle\\
&=& \sum_{(x)} \epsilon(h_{2}) \epsilon(h_{1}) f_{2}[S(x''')] f_{1}[S(x'')] |x'\rangle\\
&=& \sum_{(x)} \epsilon(h_{1}h_{2}) \langle f_{2} \otimes f_{1}, \Delta[S(x'')]\rangle |x'\rangle\\
&=& F^{(h_{1}h_{2},f_{2}f_{1})}(\tau_L) |x\rangle
\end{eqnarray*}
\end{tabular}
\end{center}

\begin{center}
\begin{tabular}{m{4cm} | m{7cm}}
\begin{center}
\begin{tikzpicture}
\draw[black, thick] (0,0)--(-1,1)--(1,1)--cycle;
\draw[->, black, thick] (-1,1)--(0,1) node[anchor=south] {$x$};
\draw[double, ->, >=stealth, black] (-0.3,0.65)--(0.3,0.65);
\filldraw[red] (0,0) circle (2pt);
\draw[red, very thick] (-1,1)--(0,0);
\draw[->, gray, thick] (-0.2,0.2) arc (135:-135:0.282843);
\end{tikzpicture}
\end{center}
&
\begin{eqnarray*}
&\;& F^{(h_{1},f_{1})}(\tau_L) F^{(h_{2},f_{2})}(\tau_L) |x\rangle\\ 
&=& \sum_{(x)} F^{(h_{1},f_{1})}(\tau_L) \epsilon(h_{2})f_{2}(x')|x''\rangle\\
&=& \sum_{(x)} \epsilon(h_{2}) \epsilon(h_{1}) f_{2}(x') f_{1}(x'') |x'''\rangle\\
&=& \sum_{(x)} \epsilon(h_{1}h_{2}) \langle f_{2} \otimes f_{1}, \Delta(x')\rangle |x''\rangle\\
&=& F^{(h_{1}h_{2},f_{2}f_{1})}(\tau_L) |x\rangle
\end{eqnarray*}
\end{tabular}
\end{center}

\begin{center}
\begin{tabular}{m{4cm} | m{7cm}}
\begin{center}
\begin{tikzpicture}
\draw[black, thick] (0,0)--(1,1)--(2,0);
\draw[black, dashed, thick] (0,0)--(2,0);
\draw[black, thick] (1,-1)--(1,1);
\draw[->, black, thick] (1,-1)--(1,-0.5) node[anchor=west] {$x$};
\draw[double, ->, >=stealth, black] (0.7,0.35)--(1.3,0.35);
\filldraw[red] (0,0) circle (2pt);
\draw[red, very thick] (0,0)--(1,1);
\draw[->, gray, thick] (0.2,0.2) arc (45:-225:0.282843);
\end{tikzpicture}
\end{center}
&
\begin{eqnarray*}
&\;& F^{(h_{1},f_{1})}(\tau_L) F^{(h_{2},f_{2})}(\tau_L) |x\rangle\\ 
&=& F^{(h_{1},f_{1})}(\tau_L) \epsilon(f_{2}) |xS(h_{2})\rangle\\
&=& \epsilon(f_{2}) \epsilon(f_{1}) |xS(h_{2})S(h_{1})\rangle\\
&=& \epsilon(f_{2}f_{1}) |xS(h_{1}h_{2})\rangle\\
&=& F^{(h_{1}h_{2},f_{2}f_{1})}(\tau_L) |x\rangle
\end{eqnarray*}
\end{tabular}
\end{center}

\begin{center}
\begin{tabular}{m{4cm} | m{7cm}}
\begin{center}
\begin{tikzpicture}
\draw[black, thick] (0,0)--(1,1)--(2,0);
\draw[black, dashed, thick] (0,0)--(2,0);
\draw[black, thick] (1,-1)--(1,1);
\draw[->, black, thick] (1,0)--(1,-0.5) node[anchor=west] {$x$};
\draw[double, ->, >=stealth, black] (0.7,0.35)--(1.3,0.35);
\filldraw[red] (0,0) circle (2pt);
\draw[red, very thick] (0,0)--(1,1);
\draw[->, gray, thick] (0.2,0.2) arc (45:-225:0.282843);
\end{tikzpicture}
\end{center}
&
\begin{eqnarray*}
&\;& F^{(h_{1},f_{1})}(\tau_L) F^{(h_{2},f_{2})}(\tau_L) |x\rangle\\ 
&=& F^{(h_{1},f_{1})}(\tau_L) \epsilon(f_{2}) |h_{2}x\rangle\\
&=& \epsilon(f_{2}) \epsilon(f_{1}) |h_{1}h_{2}x\rangle\\
&=& F^{(h_{1}h_{2},f_{2}f_{1})}(\tau_L) |x\rangle
\end{eqnarray*}
\end{tabular}
\end{center}

\subsection{For locally counterclockwise ribbons  $\tau_R$}
\begin{center}
\begin{tabular}{m{4cm} | m{7cm}}
\begin{center}
\begin{tikzpicture}
\draw[black, thick] (0,0)--(1,1)--(2,0)--cycle;
\draw[->, black, thick] (0,0)--(1,0) node[anchor=north] {$x$};
\draw[double, ->, >=stealth, black] (0.7,0.35)--(1.3,0.35);
\filldraw[red] (1,1) circle (2pt);
\draw[red, very thick] (0,0)--(1,1);
\draw[->, gray, thick] (0.8,0.8) arc (-135:135:0.282843);
\end{tikzpicture}
\end{center}
&
\begin{eqnarray*}
&\;& F^{(h_{1},f_{1})}(\tau_R) F^{(h_{2},f_{2})}(\tau_R) |x\rangle\\ 
&=& \sum_{(x)} F^{(h_{1},f_{1})}(\tau_R) \epsilon(h_{2})f_{2}(x'')|x'\rangle\\
&=& \sum_{(x)} \epsilon(h_{2}) \epsilon(h_{1}) f_{2}(x''') f_{1}(x'') |x'\rangle\\
&=& \sum_{(x)} \epsilon(h_{2}h_{1}) \langle f_{1} \otimes f_{2}, \Delta(x'')\rangle |x'\rangle\\
&=& F^{(h_{2}h_{1},f_{1}f_{2})}(\tau_R) |x\rangle
\end{eqnarray*}
\end{tabular}
\end{center}

\begin{center}
\begin{tabular}{m{4cm} | m{7cm}} 
\begin{center}
\begin{tikzpicture}
\draw[black, thick] (0,0)--(1,1)--(2,0)--cycle;
\draw[->, black, thick] (2,0)--(1,0) node[anchor=north] {$x$};
\draw[double, ->, >=stealth, black] (0.7,0.35)--(1.3,0.35);
\filldraw[red] (1,1) circle (2pt);
\draw[red, very thick] (0,0)--(1,1);
\draw[->, gray, thick] (0.8,0.8) arc (-135:135:0.282843);
\end{tikzpicture}
\end{center}
&
\begin{eqnarray*}
&\;& F^{(h_{1},f_{1})}(\tau_R) F^{(h_{2},f_{2})}(\tau_R) |x\rangle\\ 
&=& \sum_{(x)} F^{(h_{1},f_{1})}(\tau_R) \epsilon(h_{2})f_{2}[S(x')]|x''\rangle\\
&=& \sum_{(x)} \epsilon(h_{2}) \epsilon(h_{1}) f_{2}[S(x')] f_{1}[S(x'')]|x'''\rangle\\
&=& \sum_{(x)} \epsilon(h_{2}h_{1}) \langle f_{1} \otimes f_{2}, \Delta[S(x')]\rangle |x''\rangle\\
&=& F^{(h_{2}h_{1},f_{1}f_{2})}(\tau_R) |x\rangle
\end{eqnarray*}
\end{tabular}
\end{center}

\begin{center}
\begin{tabular}{m{4cm} | m{7cm}}
\begin{center}
\begin{tikzpicture}
\draw[black, thick] (-1,1)--(0,0)--(1,1);
\draw[black, dashed, thick] (-1,1)--(1,1);
\draw[black, thick] (0,2)--(0,0);
\draw[->, black, thick] (0,1)--(0,1.5) node[anchor=west] {$x$};
\draw[double, ->, >=stealth, black] (-0.3,0.65)--(0.3,0.65);
\filldraw[red] (-1,1) circle (2pt);
\draw[red, very thick] (-1,1)--(0,0);
\draw[->, gray, thick] (-0.8,0.8) arc (-45:225:0.282843);
\end{tikzpicture}
\end{center}
&
\begin{eqnarray*}
&\;& F^{(h_{1},f_{1})}(\tau_R) F^{(h_{2},f_{2})}(\tau_R) |x\rangle\\ 
&=& F^{(h_{1},f_{1})}(\tau_R) \epsilon(f_{2}) |S(h_{2})x\rangle\\
&=& \epsilon(f_{2}) \epsilon(f_{1}) |S(h_{1})S(h_{2})x\rangle\\
&=& \epsilon(f_{1}f_{2}) |S(h_{2}h_{1})x\rangle\\
&=& F^{(h_{2}h_{1},f_{1}f_{2})}(\tau_R) |x\rangle
\end{eqnarray*}
\end{tabular}
\end{center}

\begin{center}
\begin{tabular}{m{4cm} | m{7cm}}
\begin{center}
\begin{tikzpicture}
\draw[black, thick] (-1,1)--(0,0)--(1,1);
\draw[black, dashed, thick] (-1,1)--(1,1);
\draw[black, thick] (0,2)--(0,0);
\draw[->, black, thick] (0,2)--(0,1.5) node[anchor=west] {$x$};
\draw[double, ->, >=stealth, black] (-0.3,0.65)--(0.3,0.65);
\filldraw[red] (-1,1) circle (2pt);
\draw[red, very thick] (-1,1)--(0,0);
\draw[->, gray, thick] (-0.8,0.8) arc (-45:225:0.282843);
\end{tikzpicture}
\end{center}
&
\begin{eqnarray*}
&\;& F^{(h_{1},f_{1})}(\tau_R) F^{(h_{2},f_{2})}(\tau_R) |x\rangle\\ 
&=& F^{(h_{1},f_{1})}(\tau_R) \epsilon(f_{2}) |xh_{2}\rangle\\
&=& \epsilon(f_{2}) \epsilon(f_{1}) |xh_{2}h_{1}\rangle\\
&=& \epsilon(f_{1}f_{2}) |xh_{2}h_{1}\rangle\\
&=& F^{(h_{2}h_{1},f_{1}f_{2})}(\tau_R) |x\rangle
\end{eqnarray*}
\end{tabular}
\end{center}

\section{Proof of Lemma \ref{lem:local_operator_at_ends}}\label{sec:Local actions on ribbon operators}
The idea is to first prove the equations in Lemma \ref{lem:local_operator_at_ends} for ribbons as short as possible, and then extend them to longer ribbons. It turns out that the shortest ribbon for some of the equations to hold is a triangle (direct or dual), while for others is a 2-triangle. For example, see the ribbon in Subsection  \ref{subsec:for eqn a at s0}. Equation \ref{eqn:a at s0} does not hold for the rightmost triangle alone. This is roughly because for that triangle, its initial site and terminal site share the same vertex so that $A_a(s_0)$ would also act on $s_1$, which is unexpected. As will be shown below, the equation does hold as long as we make the triangle a bit longer. This is not a problem since we are only interested in properties of sufficiently long ribbons.

Subsections \ref{subsec:for eqn a at s0}-\ref{subsec:for eqn d at s1} each addresses an identity in Equations \ref{eqn:a at s0} - \ref{eqn:d at s1} for the shortest possible ribbon. For each of the eight equations, there are two types of triangles (direct or dual) to consider. To avoid lengthy calculations, we only present the details for one of the two types for each equation. The proof for the other cases is similar. If a triangle does not work, then we lengthen it to a 2-triangle. In Subsection \ref{subsec:long ribbon for s0} we extend the results to longer ribbons for Equations \ref{eqn:b at s0} and \ref{eqn:c at s0} while leave the other six cases as an exercise (whose proof is similar as well). 

\subsection{Equation \ref{eqn:a at s0} for short ribbons}
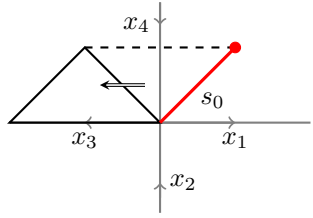
\label{subsec:for eqn a at s0}
\begin{center}
\begin{tikzpicture}
\draw[gray, thick] (0,-1.2)--(0,1.6);
\draw[gray, thick] (-2,0)--(2,0);
\draw[->, gray, thick] (0,0)--(1,0) node[anchor=north,black] {$x_1$};
\draw[->, gray, thick] (0,-1)--(0,-0.8) node[anchor=west,black] {$x_2$};
\draw[->, gray, thick] (0,0)--(-1,0) node[anchor=north,black] {$x_3$};
\draw[->, gray, thick] (0,1.6)--(0,1.3) node[anchor=east,black] {$x_4$};
\draw[black, thick] (0,0)--(-1,1)--(-2,0)--cycle;
\draw[black, dashed, thick] (1,1)--(-1,1);
\draw[double, ->, >=stealth, black] (-0.2,0.5)--(-0.8,0.5);
\filldraw[red] (1,1) circle (2pt);
\node[black] at (0.7,0.3) {$s_0$};
\draw[red, very thick] (0,0)--(1,1);
\end{tikzpicture}
\end{center}
\begin{align*}
&\quad A_{a}(s_{0}) F^{(h,f)}(\tau_L) |x_{1} \; x_{4} \; x_{3} \; x_{2} \rangle\\
&= A_{a}(s_{0}) \sum_{(i),i,(h)} F^{(h',g_{i})}(\tau_{1}) F^{[S(i''')h''i',f(i''?)]}(\tau_{2}) |x_{1} \; x_{4} \; x_{3} \; x_{2} \rangle\\
&= A_{a}(s_{0}) \sum_{(h),(i),i,(x_3)} F^{(h',g_{i})}(\tau_{1}) \epsilon[S(i''')h''i'] f(i'' x_{3}') |x_{1} \; x_{4} \; x_{3}'' \; x_{2} \rangle\\
&= A_{a}(s_{0}) \sum_{(h),(i),i,(x_3)} \epsilon(g_{i}) \epsilon[S(i''')h''i'] f(i'' x_{3}') |x_{1} \; x_{4}S(h') \; x_{3}'' \; x_{2} \rangle\\
&= A_{a}(s_{0}) \sum_{(e),(x_3),(h)} \epsilon(e''') \epsilon(h'') \epsilon(e') f(e'' x_{3}') |x_{1} \; x_{4}S(h') \; x_{3}'' \; x_{2} \rangle\\
&= \sum_{(a),(x_3),(h)} f(e x_{3}') |a^{(4)}x_{1} \; x_{4}S(h')S(a') \; a''x_{3}'' \; x_{2}S(a''') \rangle\\
&= \sum_{(a),(x_3)} f[\epsilon(a'')x_{3}'] |a^{(5)}x_{1} \; x_{4}\epsilon(a^{(6)})S(h)S(a') \; a'''x_{3}'' \; x_{2}S(a^{(4)}) \rangle\\
&= \sum_{(a),(x_3)} f[S(a'')e a'''x_{3}'] |a^{(6)}x_{1} \; x_{4}S(h)S(a') \; a^{(4)}x_{3}'' \; x_{2}S(a^{(5)}) \rangle\\
&= \sum_{(i),i,(h),(x_3),(a)} \epsilon(g_{i}) \epsilon(i''') \epsilon\{[a''h''S(a^{(4)})]\} \epsilon(i')f[S(a''')i'' a^{(7)} x_{3}']\\
&\qquad \qquad |a^{(10)}x_{1} \; x_{4}S(a^{(6)})S\{[a'h'S(a^{(5)})]\} \; a^{(8)}x_{3}'' \; x_{2}S(a^{(9)}) \rangle\\
&= \sum_{(i),i,(x_3),(a),(x_3)} F^{\{[a'hS(a''')]',g_{i}\}}(\tau_{1}) \epsilon\{S(i''')[a'hS(a''')]''i'\} f[S(a'')i'' (a''')'x_{3}']\\
&\qquad \qquad |a^{(7)}x_{1} \; x_{4}S(a^{(4)}) \; (a^{(5)})''x_{3}'' \; x_{2}S(a^{(6)}) \rangle\\
&= \sum_{(i),i,(a),(a'hS(a'''))} F^{\{[a'hS(a''')]',g_{i}\}}(\tau_{1}) F^{\{S(i''')[a'hS(a^{(3)})]''i',f[S(a'')i'' ?]\}}(\tau_{2})\\
&\qquad \qquad |a^{(7)}x_{1} \; x_{4}S(a^{(4)}) \; a^{(5)}x_{3} \; x_{2}S(a^{(6)}) \rangle\\
&= \sum_{(i),(a)} F^{\{a'hS(a'''),f[S(a'')?]\}}(\tau_L) |a^{(7)}x_{1} \; x_{4}S(a^{(4)}) \; a^{(5)}x_{3} \; x_{2}S(a^{(6)}) \rangle\\
&= \sum_{(i),(a)} F^{\{a'hS(a^{(3)}),f[S(a'')?]\}}(\tau_L) A_{a^{(4)}}(s_{0}) |x_{1} \; x_{4} \; x_{3} \; x_{2} \rangle
\end{align*}

From the fourth line to the fifth line above, we used $\epsilon(g_i)=g_{i}(e)$ and
\begin{equation*}
\sum_{(i),i} g_{i}(a)i'f(i'')= \sum_{(a)} a'f(a''). 
\end{equation*}
To derive the above equality, note that,
\begin{equation*}
    \sum_{(a)} a'f(a'')=(Id \otimes f)\Delta(a) =(Id \otimes f)\Delta\left(\sum_{i} g_i(a) i\right).
\end{equation*}

\subsection{Equation \ref{eqn:b at s0}  for short ribbons}
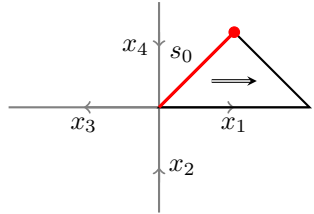
\begin{center}
\begin{tikzpicture}
\draw[gray, thick] (0,-1.4)--(0,1.4);
\draw[gray, thick] (-2,0)--(2,0);
\draw[->, gray, thick] (0,0)--(1,0) node[anchor=north,black] {$x_1$};
\draw[->, gray, thick] (0,-1)--(0,-0.8) node[anchor=west,black] {$x_2$};
\draw[->, gray, thick] (0,0)--(-1,0) node[anchor=north,black] {$x_3$};
\draw[->, gray, thick] (0,1)--(0,0.8) node[anchor=east,black] {$x_4$};
\draw[black, thick] (0,0)--(1,1)--(2,0)--cycle;
\draw[double, ->, >=stealth, black] (0.7,0.35)--(1.3,0.35);
\filldraw[red] (1,1) circle (2pt);
\node[black] at (0.3,0.7) {$s_0$};
\draw[red, very thick] (0,0)--(1,1);
\end{tikzpicture}
\end{center}
\begin{align*}
&\quad A_{a}(s_{0}) F^{(h,f)}(\tau_R) |x_{1} \; x_{4} \; x_{3} \; x_{2} \rangle\\
&= A_{a}(s_{0}) \sum_{(x_1)} \epsilon(h) f(x_{1}'') |x_{1}' \; x_{4} \; x_{3} \; x_{2} \rangle\\
&= \sum_{(x_1),(a)} \epsilon(h) f(x_{1}'') |a^{(4)}x_{1}' \; x_{4}S(a') \; a''x_{3} \; x_{2}S(a''') \rangle\\
&= \sum_{(x_1),(a)} \epsilon(a^{(6)}) \epsilon(h) \epsilon(a^{(8)}) f[S(a^{(7)})a^{(5)}x_{1}''] |a^{(4)}x_{1}' \; x_{4}S(a') \; a''x_{3} \; x_{2}S(a''') \rangle\\
&= \sum_{(x_1),(a)} \epsilon[a^{(5)}hS(a^{(7)})] f[S(a^{(6)})(a^{(4)}x_{1})''] |(a^{(4)}x_{1})' \; x_{4}S(a') \; a''x_{3} \; x_{2}S(a''') \rangle\\
&= \sum_{(a)} F^{\{a^{(5)}hS(a^{(7)}), f[S(a^{(6)})?]\}}(\tau_R) |a^{(4)}x_{1} \; x_{4}S(a') \; a''x_{3} \; x_{2}S(a''') \rangle\\
&= \sum_{(a)} F^{\{a''hS(a^{(4)}), f[S(a''')?]\}}(\tau_R)A_{a'}(s_{0}) |x_{1} \; x_{4} \; x_{3} \; x_{2} \rangle
\end{align*}

\subsection{Equation \ref{eqn:c at s0}  for short ribbons}
\begin{center}
\begin{tikzpicture}
\draw[gray, thick] (-1,-1) rectangle (1,1);
\draw[gray, thick] (-1,1)--(1,-1);
\draw[gray, thick] (-1,-1)--(1,1);
\draw[->, gray, thick] (-1,-1)--(0,-1) node[anchor=north,black] {$x_1$};
\draw[->, gray, thick] (1,0)--(1,0.5) node[anchor=west,black] {$x_2$};
\draw[->, gray, thick] (1,1)--(0,1) node[anchor=south,black] {$x_3$};
\draw[->, gray, thick] (-1,1)--(-1,0.5) node[anchor=east,black] {$x_4$};
\draw[black, thick] (0,0)--(-1,-1)--(-2,0);
\draw[black, dashed, thick] (0,0)--(-2,0);
\draw[double, ->, >=stealth, black] (-0.7,-0.35)--(-1.3,-0.35);
\filldraw[red] (0,0) circle (2pt);
\node[black] at (-0.3,-0.7) {$s_0$};
\draw[red, very thick] (-1,-1)--(0,0);
\end{tikzpicture}
\end{center}
\begin{align*}
&\quad B_{t}(s_{0}) F^{(h,f)}(\tau_L) |x_{1} \; x_{2} \; x_{3} \; x_{4} \rangle\\
&= B_{t}(s_{0}) \epsilon(f) |x_{1} \; x_{2} \; x_{3} \; x_{4}S(h) \rangle\\
&= \sum_{(x_i),(h)} \epsilon(f) t[x_{1}''x_{2}''x_{3}''x_{4}''S(h')] |x_{1}' \; x_{2}' \; x_{3}' \; x_{4}'S(h'') \rangle\\
&= \sum_{(x_i),(h)} F^{(h'',f)}(\tau_L) t[x_{1}''x_{2}''x_{3}''x_{4}''S(h')] |x_{1}' \; x_{2}' \; x_{3}' \; x_{4}' \rangle\\
&= \sum_{(h)} F^{(h'',f)}(\tau_L) B_{t[?S(h')]}(s_{0}) |x_{1} \; x_{2} \; x_{3} \; x_{4} \rangle
\end{align*}

\subsection{Equation \ref{eqn:d at s0}  for short ribbons}
\begin{center}
\begin{tikzpicture}
\draw[gray, thick] (-1,-1) rectangle (1,1);
\draw[gray, thick] (-1,1)--(1,-1);
\draw[gray, thick] (-1,-1)--(1,1);
\draw[->, gray, thick] (-1,-1)--(0,-1) node[anchor=north,black] {$x_1$};
\draw[->, gray, thick] (1,0)--(1,0.5) node[anchor=west,black] {$x_2$};
\draw[->, gray, thick] (1,1)--(0,1) node[anchor=south,black] {$x_3$};
\draw[->, gray, thick] (-1,1)--(-1,0.5) node[anchor=east,black] {$x_4$};
\draw[black, thick] (0,0)--(1,-1)--(2,0);
\draw[black, dashed, thick] (0,0)--(2,0);
\draw[black, thick] (-1,-1)--(1,-1);
\draw[double, ->, >=stealth, black] (0.2,-0.5)--(0.8,-0.5);
\filldraw[red] (0,0) circle (2pt);
\node[black] at (-0.7,-0.3) {$s_0$};
\draw[red, very thick] (-1,-1)--(0,0);
\end{tikzpicture}
\end{center}

\begin{align*}
&\quad B_{t}(s_{0}) F^{(h,f)}(\tau_R) |x_{1} \; x_{2} \; x_{3} \; x_{4} \rangle\\
&= B_{t}(s_{0}) \sum_{(h),i,(i)} F^{(h',g_{i})}(\tau_{1}) F^{[S(i''')h''i',f(i'' ?)]}(\tau_{2}) |x_{1} \; x_{2} \; x_{3} \; x_{4} \rangle\\
&= B_{t}(s_{0}) \sum_{(h),i,(i)} F^{(h',g_{i})}(\tau_{1}) \epsilon[f(i'' ?)] |x_{1} \; S[S(i''')h''i']x_{2} \; x_{3} \; x_{4} \rangle\\
&= B_{t}(s_{0}) \sum_{(h),i,(i),(x_1)} \epsilon(h') g_{i}(x_{1}'') f(i'') |x_{1}' \; S(i')S(h'')i'''x_{2} \; x_{3} \; x_{4} \rangle\\
&= B_{t}(s_{0}) \sum_{(x_1)} f(x_{1}''') |x_{1}' \; S(x_{1}'')S(h)x_{1}^{(4)}x_{2} \; x_{3} \; x_{4} \rangle\\
&= \sum_{(h),(x_i)} f(x_{1}^{(5)}) t[x_{1}''S(x_{1}''')S(h')x_{1}^{(7)}x_{2}''x_{3}''x_{4}''] |x_{1}' \; S(x_{1}^{(4)})S(h'')x_{1}^{(6)}x_{2}' \; x_{3}' \; x_{4}' \rangle\\
&= \sum_{(h),(x_i)} f(x_{1}^{(4)}) t[\epsilon(x_{1}'')S(h')x_{1}^{(6)}x_{2}''x_{3}''x_{4}''] |x_{1}' \; S(x_{1}''')S(h'')x_{1}^{(5)}x_{2}' \; x_{3}' \; x_{4}' \rangle\\
&= \sum_{(h),(x_i)} f(x_{1}''') t[S(h)x_{1}^{(5)}x_{2}''x_{3}''x_{4}''] |x_{1}' \; S(x_{1}'')S(h'')x_{1}^{(4)}x_{2}' \; x_{3}' \; x_{4}' \rangle\\
&= \sum_{(h),(x_i),i,(i)} \epsilon(h'') g_{i}(x_{1}'')f(i'') t[S(h')x_{1}''x_{2}''x_{3}''x_{4}''] |x_{1}' \; S(i')S(h''')i'''x_{2}' \; x_{3}' \; x_{4}' \rangle\\
&= \sum_{(h),(x_i),i,(i)} F^{(h'',g_{i})}(\tau_{1}) \epsilon[f(i'' ?)] t[S(h')x_{1}''x_{2}''x_{3}''x_{4}''] |x_{1}' \; S[S(i''')h'''i']x_{2}' \; x_{3}' \; x_{4}' \rangle\\
&= \sum_{(h),(x_i),i,(i)} F^{(h'',g_{i})}(\tau_{1}) F^{[S(i''')h'''i',f(i'' ?)]}(\tau_{2}) t[S(h')x_{1}''x_{2}''x_{3}''x_{4}''] |x_{1}' \; x_{2}' \; x_{3}' \; x_{4}' \rangle\\
&= \sum_{(h),(x_i)} F^{(h'',f)}(\tau_R) t[S(h')x_{1}''x_{2}''x_{3}''x_{4}''] |x_{1}' \; x_{2}' \; x_{3}' \; x_{4}' \rangle\\
&= \sum_{(h)} F^{(h'',f)}(\tau_R) B_{t[S(h')?]}(s_{0}) |x_{1} \; x_{2} \; x_{3} \; x_{4} \rangle
\end{align*}

\subsection{Equation \ref{eqn:a at s1}  for short ribbons}
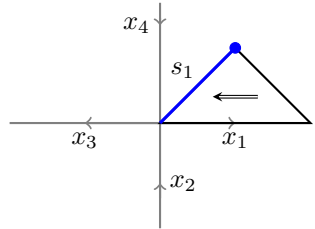
\begin{center}
\begin{tikzpicture}
\draw[gray, thick] (0,-1.4)--(0,1.4);
\draw[gray, thick] (-2,0)--(2,0);
\draw[->, gray, thick] (0,0)--(1,0) node[anchor=north,black] {$x_1$};
\draw[->, gray, thick] (0,-1)--(0,-0.8) node[anchor=west,black] {$x_2$};
\draw[->, gray, thick] (0,0)--(-1,0) node[anchor=north,black] {$x_3$};
\draw[->, gray, thick] (0,1.6)--(0,1.3) node[anchor=east,black] {$x_4$};
\draw[black, thick] (0,0)--(1,1)--(2,0)--cycle;
\draw[double, ->, >=stealth, black] (1.3,0.35)--(0.7,0.35);
\filldraw[blue] (1,1) circle (2pt);
\node[black] at (0.3,0.7) {$s_1$};
\draw[blue, very thick] (0,0)--(1,1);
\end{tikzpicture}
\end{center}
\begin{align*}
&\quad A_{a}(s_{1}) F^{(h,f)}(\tau_L) |x_{1} \; x_{4} \; x_{3} \; x_{2} \rangle\\
&= A_{a}(s_{1}) \sum_{(x_1)} \epsilon(h) f(x_{1}'') |x_{1}' \; x_{4} \; x_{3} \; x_{2} \rangle\\
&= \sum_{(x_1),(a)} \epsilon(h) f[S(x_{1}'')] |a^{(4)}x_{1}' \; x_{4}S(a') \; a''x_{3} \; x_{2}S(a''') \rangle\\
&= \sum_{(x_1),(a)} \epsilon(h) f[S(x_{1}'') \epsilon(a^{(5)})] |a^{(4)}x_{1}' \; x_{4}S(a') \; a''x_{3} \; x_{2}S(a''') \rangle\\
&= \sum_{(x_1),(a)} \epsilon(h) f[S(x_{1}'')S(a^{(5)})a^{(6)}] |a^{(4)}x_{1}' \; x_{4}S(a') \; a''x_{3} \; x_{2}S(a''') \rangle\\
&= \sum_{(a)} F^{[h, f(?a^{(5)})]}(\tau_L) |a^{(4)}x_{1} \; x_{4}S(a') \; a''x_{3} \; x_{2}S(a''') \rangle\\
&= \sum_{(a)} F^{[h, f(?a'')]}(\tau_L)A_{a'}(s_{1}) |x_{1} \; x_{4} \; x_{3} \; x_{2} \rangle
\end{align*}

\subsection{Equation \ref{eqn:b at s1}  for short ribbons}
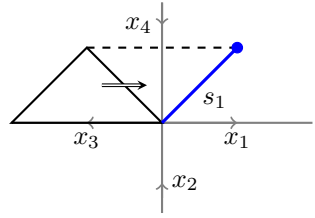
\begin{center}
\begin{tikzpicture}
\draw[gray, thick] (0,-1.2)--(0,1.6);
\draw[gray, thick] (-2,0)--(2,0);
\draw[->, gray, thick] (0,0)--(1,0) node[anchor=north,black] {$x_1$};
\draw[->, gray, thick] (0,-1)--(0,-0.8) node[anchor=west,black] {$x_2$};
\draw[->, gray, thick] (0,0)--(-1,0) node[anchor=north,black] {$x_3$};
\draw[->, gray, thick] (0,1.6)--(0,1.3) node[anchor=east,black] {$x_4$};
\draw[black, thick] (0,0)--(-1,1)--(-2,0)--cycle;
\draw[black, dashed, thick] (1,1)--(-1,1);
\draw[double, ->, >=stealth, black] (-0.8,0.5)--(-0.2,0.5);
\filldraw[blue] (1,1) circle (2pt);
\node[black] at (0.7,0.3) {$s_1$};
\draw[blue, very thick] (0,0)--(1,1);
\end{tikzpicture}
\end{center}
\begin{align*}
&\quad A_{a}(s_{1}) F^{(h,f)}(\tau_R) |x_{1} \; x_{4} \; x_{3} \; x_{2} \rangle\\
&= A_{a}(s_{1}) \sum_{(i),i,(h)} F^{(h',g_{i})}(\tau_{1}) F^{[S(i''')h''i',f(i'' ?)]}(\tau_{2}) |x_{1} \; x_{4} \; x_{3} \; x_{2} \rangle\\
&= A_{a}(s_{1}) \sum_{(i),i,(h)} F^{(h',g_{i})}(\tau_{1}) \epsilon[f(i' ?)] |x_{1} \; x_{4}S(i''')h''i' \; x_{3} \; x_{2} \rangle\\
&= A_{a}(s_{1}) \sum_{(i),i,(h),(x_3)} \epsilon(h') g_{i}[S(x_{3}')] f(i'') |x_{1} \; x_{4}S(i''')h''i' \; x_{3}'' \; x_{2} \rangle\\
&= \sum_{(x_3)} A_{a}(s_{1}) f[S(x_{3}'')] |x_{1} \; x_{4}x_{3}'hS(x_{3}''') \; x_{3}^{(4)} \; x_{2} \rangle\\
&= \sum_{(a),(x_3)} f[S(x_{3}'')] |a^{(4)}x_{1} \; x_{4}x_{3}'hS(x_{3}''')S(a') \; a''x_{3}^{(4)} \; x_{2}S(a''') \rangle\\
&= \sum_{(a),(x_3)} f[S(x_{3}'')S(a'')a'] |a^{(6)}x_{1} \; x_{4}x_{3}'hS(x_{3}''')S(a''') \; a^{(4)}x_{3}^{(4)} \; x_{2}S(a^{(5)}) \rangle\\
&= \sum_{(a),(x_3)} f[S(a^{(4)}x_{3}'') a'] |a^{(8)}x_{1} \; x_{4}S(a'')a'''x_{3}'hS(x_{3}''')S(a^{(5)}) \; a^{(6)}x_{3}^{(4)} \; x_{2}S(a^{(7)}) \rangle\\
&= \sum_{(a),(i),i,(h),(x_3)} \epsilon(h') g_{i}[S(a'''x_{3}')] \epsilon[f(i'' ?a')]\\
&\qquad \qquad |a^{(6)}x_{1} \; x_{4}S(a'')S(i''')h''i' \; a^{(4)}x_{3}'' \; x_{2}S(a^{(5)}) \rangle\\
&= \sum_{(a),(i),i,(h)} F^{(h',g_{i})}(\tau_{1}) \epsilon[f(i'' ?a')] |a^{(5)}x_{1} \; x_{4}S(a'')S(i''')h''i' \; a'''x_{3} \; x_{2}S(a^{(4)}) \rangle\\
&= \sum_{(a),(i),i,(h)} F^{(h',g_{i})}(\tau_{1}) F^{[S(i''')h''i',f(i'' ?a')]}(\tau_{2}) |a^{(5)}x_{1} \; x_{4}S(a'') \; a'''x_{3} \; x_{2}S(a^{(4)}) \rangle\\
&= \sum_{(a)} F^{[h,f(?a')]}(\tau_R) |a^{(5)}x_{1} \; x_{4}S(a'') \; a'''x_{3} \; x_{2}S(a^{(4)}) \rangle\\
&= \sum_{(a)} F^{[h,f(?a')]}(\tau_R) A_{a''}(s_{1}) |x_{1} \; x_{4} \; x_{3} \; x_{2} \rangle
\end{align*}

\subsection{Equation \ref{eqn:c at s1}  for short ribbons}
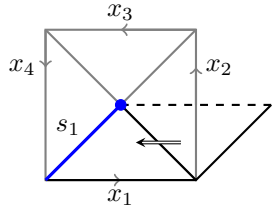
\begin{center}
\begin{tikzpicture}
\draw[gray, thick] (-1,-1) rectangle (1,1);
\draw[gray, thick] (-1,1)--(1,-1);
\draw[gray, thick] (-1,-1)--(1,1);
\draw[->, gray, thick] (-1,-1)--(0,-1) node[anchor=north,black] {$x_1$};
\draw[->, gray, thick] (1,0)--(1,0.5) node[anchor=west,black] {$x_2$};
\draw[->, gray, thick] (1,1)--(0,1) node[anchor=south,black] {$x_3$};
\draw[->, gray, thick] (-1,1)--(-1,0.5) node[anchor=east,black] {$x_4$};
\draw[black, thick] (0,0)--(1,-1)--(2,0);
\draw[black, dashed, thick] (0,0)--(2,0);
\draw[black, thick] (-1,-1)--(1,-1);
\draw[double, ->, >=stealth, black] (0.8,-0.5)--(0.2,-0.5);
\filldraw[blue] (0,0) circle (2pt);
\node[black] at (-0.7,-0.3) {$s_1$};
\draw[blue, very thick] (-1,-1)--(0,0);
\end{tikzpicture}
\end{center}
\begin{align*}
&\quad B_{t}(s_{1}) F^{(h,f)}(\tau_L) |x_{1} \; x_{2} \; x_{3} \; x_{4} \rangle\\
&= B_{t}(s_{1}) \sum_{(h),i,(i)} F^{(h',g_{i})}(\tau_{1}) F^{[S(i''')h''i',f(i'' ?)]}(\tau_{2}) |x_{1} \; x_{2} \; x_{3} \; x_{4} \rangle\\
&= B_{t}(s_{1}) \sum_{(h),i,(i),(x_1)} F^{(h',g_{i})}(\tau_{1}) \epsilon[S(i''')h''i']f(i'' x_{1}') |x_{1}'' \; x_{2} \; x_{3} \; x_{4} \rangle\\
&= B_{t}(s_{1}) \sum_{(h),i,(i),(x_1)} \epsilon(g_{i}) \epsilon(i''') \epsilon(h'') \epsilon(i') f(i'' x_{1}') |x_{1}'' \; x_{2}S(h') \; x_{3} \; x_{4} \rangle\\
&= \sum_{(x_1)} B_{t}(s_{1}) f(x_{1}') |x_{1}'' \; x_{2}S(h) \; x_{3} \; x_{4}) \rangle\\
&= \sum_{(x_i),(h)} f(x_{1}') t[S(x_{1}'')h''S(x_{2}')S(x_{3}')S(x_{4}')] |x_{1}''' \; x_{2}''S(h') \; x_{3}'' \; x_{4}'' \rangle\\
&= \sum_{(x_i),(h)} f(x_{1}''') t[S(x_{1}^{(4)})h''x_{1}''S(x_{1}')S(x_{2}')S(x_{3}')S(x_{4}')] |x_{1}^{(5)} \; x_{2}''S(h') \; x_{3}'' \; x_{4}'' \rangle\\
&= \sum_{(x_i),i,(i),(h)} f(i'') g_{i}(x_{1}'') t[S(i''')h''i'S(x_{1}')S(x_{2}')S(x_{3}')S(x_{4}')] |x_{1}''' \; x_{2}''S(h') \; x_{3}'' \; x_{4}'' \rangle\\
&= \sum_{(x_i),i,(i),j,(j),(h)} f(i'') \epsilon(g_{i}) \epsilon(j''') \epsilon(h'') \epsilon(j') g_{i}(j'' x_{1}'')\\
&\qquad \qquad t[S(i''')h''i'S(x_{1}')S(x_{2}')S(x_{3}')S(x_{4}')] |x_{1}''' \; x_{2}''S(h') \; x_{3}'' \; x_{4}'' \rangle\\
&= \sum_{(x_i),i,(i),j,(j),(h)} f(i'') F^{(h',g_{i})}(\tau_{1}) \epsilon[S(j''')h''j'] g_{i}(j'' x_{1}'')\\
&\qquad \qquad t[S(i''')h''i'S(x_{1}')S(x_{2}')S(x_{3}')S(x_{4}')] |x_{1}''' \; x_{2}'' \; x_{3}'' \; x_{4}'' \rangle\\
&= \sum_{(x_i),i,(i),j,(j),(h)} f(i'') F^{(h',g_{i})}(\tau_{1}) F^{[S(j''')h''j',g_{i}(j'' ?)]}(\tau_{2})\\
&\qquad \qquad t[S(i''')h''i'S(x_{1}')S(x_{2}')S(x_{3}')S(x_{4}')] |x_{1}'' \; x_{2}'' \; x_{3}'' \; x_{4}'' \rangle\\
&= \sum_{(x_i),i,(i),(h)} f(i'') F^{(h',g_{i})}(\tau_L) t[S(i''')h''i'S(x_{1}')S(x_{2}')S(x_{3}')S(x_{4}')] |x_{1}'' \; x_{2}'' \; x_{3}'' \; x_{4}'' \rangle\\
&= \sum_{i,(i),(h)} f(i'') F^{(h',g_{i})}(\tau_L) B_{t[S(i''')h''i'?]}(s_{1}) |x_{1} \; x_{2} \; x_{3} \; x_{4} \rangle
\end{align*}

\subsection{Equation \ref{eqn:d at s1}  for short ribbons} \label{subsec:for eqn d at s1}
\begin{center}
\begin{tikzpicture}
\draw[gray, thick] (-1,-1) rectangle (1,1);
\draw[gray, thick] (-1,1)--(1,-1);
\draw[gray, thick] (-1,-1)--(1,1);
\draw[->, gray, thick] (-1,-1)--(0,-1) node[anchor=north,black] {$x_1$};
\draw[->, gray, thick] (1,0)--(1,0.5) node[anchor=west,black] {$x_2$};
\draw[->, gray, thick] (1,1)--(0,1) node[anchor=south,black] {$x_3$};
\draw[->, gray, thick] (-1,1)--(-1,0.5) node[anchor=east,black] {$x_4$};
\draw[black, thick] (0,0)--(-1,-1)--(-2,0);
\draw[black, dashed, thick] (0,0)--(-2,0);
\draw[double, ->, >=stealth, black] (-1.3,-0.35)--(-0.7,-0.35);
\filldraw[blue] (0,0) circle (2pt);
\node[black] at (-0.3,-0.7) {$s_1$};
\draw[blue, very thick] (-1,-1)--(0,0);
\end{tikzpicture}
\end{center}

\begin{align*}
&\quad B_{t}(s_{1}) F^{(h,f)}(\tau_R) |x_{1} \; x_{2} \; x_{3} \; x_{4} \rangle\\
&= B_{t}(s_{1}) \epsilon(f) |x_{1} \; x_{2} \; x_{3} \; x_{4}S(h) \rangle\\
&= \sum_{(x_i),(h)} f(e) t(x_{1}''x_{2}''x_{3}''x_{4}''h'') |x_{1}' \; x_{2}' \; x_{3}' \; x_{4}'h' \rangle\\
&= \sum_{(x_i),(h),i,(i)} f(i'') \epsilon(g_{i}) t[x_{1}''x_{2}''x_{3}''x_{4}''S(i''')h''i'](s_{1}) |x_{1}' \; x_{2}' \; x_{3}' \; x_{4}'h' \rangle\\
&= \sum_{(x_i),(h),i,(i)} f(i'') F^{(h',g_{i})}(\tau_R) t[x_{1}''x_{2}''x_{3}''x_{4}''S(i''')h''i'](s_{1}) |x_{1}' \; x_{2}' \; x_{3}' \; x_{4}' \rangle\\
&= \sum_{(h),i,(i)} f(i'') F^{(h',g_{i})}(\tau_R) B_{t[?S(i''')h''i']}(s_{1}) |x_{1} \; x_{2} \; x_{3} \; x_{4} \rangle
\end{align*}

\subsection{Equations \ref{eqn:b at s0} and \ref{eqn:c at s0} for long ribbons}\label{subsec:long ribbon for s0}
\begin{tabular}{m{5.5cm} | m{5.5cm}} 
\begin{center}
\begin{tikzpicture}
\draw[gray, thick] (0,-1.4)--(0,1.4);
\draw[gray, thick] (-1.2,0)--(3,0);
\draw[->, gray, thick] (0,0)--(1,0) node[anchor=north,black] {$x_1$};
\draw[->, gray, thick] (0,-1)--(0,-0.8) node[anchor=west,black] {$x_2$};
\draw[->, gray, thick] (0,0)--(-1,0) node[anchor=north,black] {$x_3$};
\draw[->, gray, thick] (0,1)--(0,0.8) node[anchor=east,black] {$x_4$};
\draw[black, thick] (0,0)--(1,1)--(2,0)--cycle;
\draw[black, thick] (2,0)--(3,1);
\draw[black, dashed, thick] (1,1)--(3,1);
\draw[double, ->, >=stealth, black] (0.7,0.35)--(1.3,0.35) node[anchor=south east] {$\tau_1$};
\draw[double, ->, >=stealth, black] (01.7,0.65)--(2.3,0.65) node[anchor=north east] {$\tau_2$};
\node at (3,0.65) {\textbf{...}};
\filldraw[red] (1,1) circle (2pt);
\node[black] at (0.3,0.7) {$s_0$};
\draw[red, very thick] (0,0)--(1,1);
\end{tikzpicture}
\end{center}
&
\begin{center}
\begin{tikzpicture}
\draw[gray, thick] (-1,-1) rectangle (1,1);
\draw[gray, thick] (-1,1)--(1,-1);
\draw[gray, thick] (-1,-1)--(1,1);
\draw[->, gray, thick] (-1,-1)--(0,-1) node[anchor=north,black] {$x_1$};
\draw[->, gray, thick] (1,0)--(1,0.5) node[anchor=west,black] {$x_2$};
\draw[->, gray, thick] (1,1)--(0,1) node[anchor=south,black] {$x_3$};
\draw[->, gray, thick] (-1,1)--(-1,0.5) node[anchor=east,black] {$x_4$};
\draw[black, thick] (0,0)--(-1,-1)--(-2,0);
\draw[black, dashed, thick] (0,0)--(-2,0);
\draw[black, thick] (-2,0)--(-1,-1)--(-3,-1)--cycle;
\draw[double, ->, >=stealth, black] (-0.7,-0.35)--(-1.3,-0.35) node[anchor=north west] {$\tau_1$};
\draw[double, ->, >=stealth, black] (-1.7,-0.65)--(-2.3,-0.65) node[anchor=south west] {$\tau_2$};
\node at (-3,-0.65) {\textbf{...}};
\filldraw[red] (0,0) circle (2pt);
\node[black] at (-0.3,-0.7) {$s_0$};
\draw[red, very thick] (-1,-1)--(0,0);
\end{tikzpicture}
\end{center}
\end{tabular}

For the left figure above, we have,
\begin{align*}
&\quad A_{a}(s_{0}) F^{(h,f)}(\tau_R)\\
&= \sum_{(h),i,(i)} A_{a}(s_{0}) F^{(h',g_{i})}(\tau_{1}) F^{[S(i''')h''i',f(i'' ?)]}(\tau_{2})\\
&= \sum_{(a),(h),i,(i)} F^{\{a''h'S(a^{(4)}), g_{i}[S(a''')?]\}}(\tau_{1}) A_{a''}(s_{0}) F^{[S(i''')h''i',f(i'' ?)]}(\tau_{2})\\
&= \sum_{(a),(h),i,(i)} F^{\{a''h'S(a^{(4)}), g_{i}[S(a''')?]\}}(\tau_{1}) F^{[S(i''')h''i',f(i'' ?)]}(\tau_{2}) A_{a'}(s_{0})\\
&= \sum_{(a),(h),i,(i),j} F^{\{a''h'S(a^{(4)}), g_{i}[S(a''')j]g_{j}(?)\}}(\tau_{1}) F^{[S(i''')h''i',f(i'' ?)]}(\tau_{2}) A_{a''}(s_{0})\\
&= \sum_{(a),(h),j,(j)} F^{[a''h'S(a^{(6)}), g_{j}]}(\tau_{1}) F^{\{S(j''')a'''h''S(a^{(5)})j', f[S(a^{(4)})j'' ?]\}}(\tau_{2}) A_{a'}(s_{0})\\
&= \sum_{(a),(h)} F^{\{a''hS(a^{(4)}), f[S(a''')?]\}}(\tau_R) A_{a'}(s_{0})
\end{align*}

From the forth line to the fifth line in the above equation, we need to use
\begin{align*} \label{sec:important technic}
g_{i}(a\;b)= g_{i}\left[a\sum_{j} g_{j}(b)j\right] = \sum_{j} g_{i}(a\;j) g_{j}(b)\\
{}\qquad\Longrightarrow  g_{i}(a\;?) = \sum_{j} g_{i}(a\;j) g_{j}(?).
\end{align*}
For the right figure above,
\begin{align*}
&\quad B_{t}(s_{0}) F^{(h,f)}(\tau_L)\\
&= \sum_{(h),i,(i)} B_{t}(s_{0}) F^{(h',g_{i})}(\tau_{1}) F^{[S(i''')h''i',f(i'' ?)]}(\tau_{2})\\
&= \sum_{(h),i,(i)} F^{(h'',g_{i})}(\tau_{1}) B_{t[?S(h')]}(s_{0}) F^{[S(i''')h'''i',f(i'' ?)]}(\tau_{2})\\
&= \sum_{(h),i,(i)} F^{(h'',g_{i})}(\tau_{1}) F^{[S(i''')h'''i',f(i'' ?)]}(\tau_{2}) B_{t[?S(h')]}(s_{0})\\
&= \sum_{(h)} F^{(h'',f)}(\tau_L) B_{t[?S(h')]}(s_{0})
\end{align*}

\section{Proof of Proposition \ref{prop:ribbon_operator_in_middle}}\label{sec:Calculation on commutation relation with ribbon in middle}
We are going to talk about Hamiltonian terms where $a$ is the Haar integral of $H$ and $t$ is the Haar integral of $H^*$ temporarily. Notice that the Haar integral is cocomutative, and so we can cyclically rotate the components $a', a'', a'''$, etc. Below we prove the commutation relation for locally clockwise ribbons, and leave the details for locally counterclockwise ribbons to the reader.
\subsection{Equation \ref{eqn:a at middle}}
\begin{center}
\begin{tikzpicture}
\draw[gray, thick] (0,-1.2)--(0,1.6);
\draw[gray, thick] (-2,0)--(2,0);
\draw[->, gray, thick] (0,0)--(1,0) node[anchor=north,black] {$x_1$};
\draw[->, gray, thick] (0,-1)--(0,-0.8) node[anchor=west,black] {$x_2$};
\draw[->, gray, thick] (0,0)--(-1,0) node[anchor=north,black] {$x_3$};
\draw[->, gray, thick] (0,1.6)--(0,1.3) node[anchor=east,black] {$x_4$};
\draw[black, thick] (0,0)--(-1,1)--(-2,0)--cycle;
\draw[black, thick] (0,0)--(1,1)--(2,0)--cycle;
\draw[black, dashed, thick] (1,1)--(-1,1);
\filldraw[brown] (1,1) circle (2pt);
\node[black] at (0.3,0.7) {$s$};
\draw[brown, very thick] (0,0)--(1,1);
\draw[double, ->, >=stealth, black] (1.3,0.35)--(0.7,0.35) node[anchor=south west] {$\tau_{1}$};
\node at (2,0.35) {\textbf{...}};
\draw[double, ->, >=stealth, black] (-0.2,0.5)--(-0.8,0.5) node[anchor=north west] {$\tau_{2}$};
\node at (-2,0.5) {\textbf{...}};
\end{tikzpicture}
\end{center}

\begin{align*}
&\quad A_a(s)F^{h,f}(\tau_L)\\
&= \sum_{(h),i,(i)} A_{a}(s) F^{(h',g_{i})}(\tau_{1}) F^{[S(i''')h''i',f(i'' ?)]}(\tau_{2})\\
&= \sum_{(h),i,(i),(a)} F^{[h',g_{i}(? a'')]}(\tau_{1}) A_{a'}(s) F^{[S(i''')h''i',f(i'' ?)]}(\tau_{2})\\
&= \sum_{(h),i,(i),(a)} F^{[h',g_{i}(? a^{(5)})]}(\tau_{1}) F^{\{a'S(i''')h''i'S(a'''), f[i'' S(a'')?]\}}(\tau_{2})  A_{a^{(4)}}(s)\\
&= \sum_{(h),i,(i),(a),j} F^{[h',g_{i}(j a^{(5)}) g_{j}(?)]}(\tau_{1}) F^{\{a'S(i''')h''i'S(a'''), f[i'' S(a'')?]\}}(\tau_{2})  A_{a^{(4)}}(s)\\
&= \sum_{(h),(a),j,(j)} F^{(h', g_{j})}(\tau_{1}) F^{\{a'S(a^{(7)})S(j''')h''j'a^{(5)}S(a'''), f[j''a^{(6)} S(a'')?]\}}(\tau_{2})  A_{a^{(4)}}(s)\\
&= \sum_{(h),(a),j,(j)} F^{(h', g_{j})}(\tau_{1}) F^{\{S(j''')h''j'a^{(4)}S(a''), f[j''a^{(5)} S(a')?]\}}(\tau_{2})  A_{a'''}(s)\\
&= \sum_{(h),(a),j,(j)} F^{(h', g_{j})}(\tau_{1})
F^{\{S(j''')h''j'a^{(3)}S(a'), f[j''?]\}}(\tau_{2})  A_{a''}(s)\\
&= \sum_{(h),j,(j)} F^{(h', g_{j})}(\tau_{1})
F^{\{S(j''')h''j', f[j''?]\}}(\tau_{2}) A_{a}(s)\\
&=F^{h,f}(\tau_L)A_{a}(s)
\end{align*}
From the sixth line to the end in the above equation, we used the cocomutative condition of $a \in H$, the Haar integral of $H$. So we can rotate $a'$ to $a^{(n_{max})}$ and $a^{(n)}$ to $a^{(n-1)}$ for $n > 1$. After the rotation, we obtain $\epsilon(a^{(n)})$ to lower the maximum order step by step.

\subsection{Equation \ref{eqn:b at middle}}
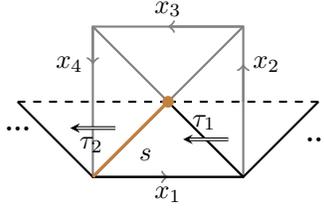
\begin{center}
\begin{tikzpicture}
\draw[gray, thick] (-1,-1) rectangle (1,1);
\draw[gray, thick] (-1,1)--(1,-1);
\draw[gray, thick] (-1,-1)--(1,1);
\draw[->, gray, thick] (-1,-1)--(0,-1) node[anchor=north,black] {$x_1$};
\draw[->, gray, thick] (1,0)--(1,0.5) node[anchor=west,black] {$x_2$};
\draw[->, gray, thick] (1,1)--(0,1) node[anchor=south,black] {$x_3$};
\draw[->, gray, thick] (-1,1)--(-1,0.5) node[anchor=east,black] {$x_4$};
\draw[black, thick] (0,0)--(-1,-1)--(-2,0);
\draw[black, dashed, thick] (0,0)--(-2,0);
\draw[black, thick] (0,0)--(1,-1)--(2,0);
\draw[black, dashed, thick] (0,0)--(2,0);
\draw[black, thick] (-1,-1)--(1,-1);
\filldraw[brown] (0,0) circle (2pt);
\node[black] at (-0.3,-0.7) {$s$};
\draw[brown, very thick] (-1,-1)--(0,0);
\draw[double, ->, >=stealth, black] (0.8,-0.5)--(0.2,-0.5) node[anchor=south west] {$\tau_{1}$};
\node at (2,-0.5) {\textbf{...}};
\draw[double, ->, >=stealth, black] (-0.7,-0.35)--(-1.3,-0.35) node[anchor=north west] {$\tau_{2}$};
\node at (-2,-0.35) {\textbf{...}};
\end{tikzpicture}
\end{center}

\begin{align*}
&\quad B_{t}(s) F^{(h,f)}(\tau_L)\\
&= \sum_{(h),i,(i)} B_{t}(s) F^{(h',g_{i})}(\tau_{1}) F^{[S(i''')h''i',f(i'' ?)]}(\tau_{2})\\
&= \sum_{(h),i,(i),j,(j)} g_{i}(j'') F^{(h',g_{j})}(\tau_{1}) B_{t[S(j''')h''j'?]}(S) F^{[S(i''')h'''i',f(i'' ?)]}(\tau_{2})\\
&= \sum_{(h),i,(i),j,(j)} g_{i}(j'') F^{(h',g_{j})}(\tau_{1}) F^{[S(i^{(4)})h^{(4)}i'',f(i''' ?)]}(\tau_{2})\\
&\qquad \qquad B_{t[S(j''')h''j'?S(i')S(h''')i^{(5)}]}(s)\\
&= \sum_{(h),j,(j)} F^{(h',g_{j})}(\tau_{1}) F^{[S(j^{(5)})h^{(4)}j''',f(j^{(4)} ?)]}(\tau_{2}) B_{t[S(j^{(7)})h''j'?S(j'')S(h''')j^{(6)}]}(s)\\
&= \sum_{(h),j,(j)} F^{(h',g_{j})}(\tau_{1}) F^{[S(j^{(5)})h^{(4)}j''',f(j^{(4)} ?)]}(\tau_{2}) B_{t[S(j'')S(h''')j^{(6)}S(j^{(7)})h''j'?]}(s)\\
&= \sum_{(h),j,(j)} F^{(h',g_{j})}(\tau_{1}) F^{[S(j''')h''j',f(j'' ?)]}(\tau_{2}) B_{t}(s)\\
&= F^{(h,f)}(\tau_L) B_{t}(s) 
\end{align*}
Similarly, from the last third line to the last second line, we used the cocomutative condition of $t \in H^*$, the Haar integral of $H^*$.

\section{Fourier transformation of $H^*$}\label{sec:FT of $H^*$}
Let $H$ be any finite dimensional $\mathbb{C}^*$ Hopf algebra. 
First, we define a Fourier transformation on $H$ \cite{buerschaper2013electric}:
\begin{equation*}
|\nu ab\rangle=\sqrt{\frac{dim(\nu)}{dim(H)}} \sum_{(h_0)} D^{\nu} (h_0')_{ab}h_0'', \quad \nu \in \text{Irr}_H, a,b = 1, \cdots, dim(\nu),
\end{equation*}
where $\text{Irr}_H$ is the set of irreducible representations of $H$, and $D^{\nu} (h_0')_{ab}$ is the matrix entry of $h_0'$ for the representation $\nu$ under a chosen (fixed) basis.

Recall from Section \ref{subsec:rep of Hopf} that there are two commuting actions, $L$ and $R$, of $H$ on itself corresponding to multiplication on the left and multiplication on the right by $S(\cdot)$, respectively. We check the form of the two actions under the Fourier basis.

For an element $m \in H$, the  action $L(m)$ is,
\begin{align*}
L(m)|\nu ab\rangle &= \sqrt{\frac{dim(\nu)}{dim(H)}} \sum_{(h_0)} D^{\nu} (h_0')_{ab}mh_0''\\
&= \sqrt{\frac{dim(\nu)}{dim(H)}} \sum_{(h_0),(m)} D^{\nu} (h_0')_{ab}m'\epsilon(m'')h_0''\\
&= \sqrt{\frac{dim(\nu)}{dim(H)}} \sum_{(h_0),(m)} D^{\nu} (h_0')_{ab}m'\epsilon[S(m'')]h_0''.
\end{align*}
As $xh_0=\epsilon(x)h_0$, we have
$\sum_{(x),(h_0)} x'h_0' \otimes x''h_0''=\epsilon(x) \sum_{(h_0)} h_0' \otimes h_0''$. Applying the above identity for $x = S(m'')$, we obtain 
\begin{align*}
L(m)|\nu ab\rangle &= \sqrt{\frac{dim(\nu)}{dim(H)}} \sum_{(h_0),(m)} D^{\nu} [S(m''')h_0']_{ab}m'S(m'')h_0''\\
&= \sqrt{\frac{dim(\nu)}{dim(H)}} \sum_{(h_0)} D^{\nu} [S(m)h_0']_{ab}h_0''\\
&= \sqrt{\frac{dim(\nu)}{dim(H)}} \sum_{(h_0),k} D^{\nu} [S(m)]_{ak} D^{\nu}(h_0')_{kb}h_0''\\
&=\sum_{k} D^{\nu} [S(m)]_{ak}|\nu kb\rangle\\
&=\sum_{k} D^{\nu^*}[m]_{ka} |\nu kb\rangle.
\end{align*}
Similarly,  we can obtain the action of $R(m)$:
\begin{equation*}
R(m)|\nu ab\rangle =\sum_{k} D^{\nu}(m)_{kb} |\nu ak\rangle.
\end{equation*}

Now, take the dual basis $\{\langle \nu a b| \}$ in $H^*$. $L$ and $R$ each induces a representation on $H^*$, still denoted by the same letter. Then on the dual basis, the two actions are given by,
\begin{align*}
L(m)(\langle \nu a b|) &= \sum_{k}D^{\nu} (m)_{ka} \langle\nu k b|, \\
R(m)(\langle \nu ab|) &= \sum_{k} D^{\nu^*}(m)_{kb}\langle \nu a k|.
\end{align*}

Apply the above dual basis to $D(H)$, we obtain the desired basis for Equations \ref{equ:dual_basis_L_act} and \ref{equ:dual_basis_R_act}.
\end{document}